\documentclass[acmsmall,screen,nonacm]{acmart}
\AtBeginDocument{%
  \providecommand\BibTeX{{%
    \normalfont B\kern-0.5em{\scshape i\kern-0.25em b}\kern-0.8em\TeX}}}

\usepackage{mathpartir}
\usepackage{xcolor}
\usepackage{cleveref}
\usepackage{graphicx}
\usepackage{caption}
\usepackage{subcaption}
\usepackage{stmaryrd}
\usepackage{bm}
\usepackage[inkscapeformat=png]{svg}
\usepackage{wrapfig}

\newtheorem{theorem}{Theorem}[section]

\newtheorem{corollary}{Corollary}[theorem]
\theoremstyle{definition}
\newtheorem{definition}{Definition}[section]

\usepackage{tikz}
\tikzstyle{box} = [
rectangle,
rounded corners,
minimum width=1cm,
minimum height=0.6cm,
text centered,
draw=black
]
\tikzstyle{label} = [
rectangle,
minimum size=1cm,
text centered,
fill=white,
draw=black
]
\tikzstyle{outer} = [
    draw=gray,dashed,thick,inner sep=5pt
]
\tikzstyle{doubblearrow} = [>=triangle 45, <->]
\tikzstyle{arrow} = [>=triangle 45, ->]

\newcommand{\cdppl}{CoreDPPL}
\newcommand{\Cdppl}{CoreDPPL}
\newcommand{\cppl}{CorePPL}
\newcommand{\miking}{Miking}
\newcommand{\lazyppl}{LazyPPL}

\newcommand{\kw}[1]{\ensuremath{\mathbf{#1}}}
\newcommand{\tyname}[1]{\ensuremath{\mathrm{#1}}}

\newcommand{\dR}{\ensuremath{\mathbb{R}}}
\newcommand{\dN}{\ensuremath{\mathbb{N}}}
\newcommand{\dVal}{\ensuremath{\mathit{Val}}}
\newcommand{\dTerm}{\ensuremath{\mathit{Term}}}
\newcommand{\dIdent}{\ensuremath{\mathit{Var}}}

\newcommand{\dom}{\ensuremath{\text{dom}}}
\newcommand{\intrinsicType}{\ensuremath{\texttt{PrimTy}}}
\newcommand{\distType}{\ensuremath{\texttt{DistTy}}}
\newcommand{\diff}[1]{\ensuremath{\texttt{Diff}#1}}
\newcommand{\solve}[1]{\ensuremath{\texttt{Solve}#1}}
\newcommand{\minfer}[1]{\ensuremath{\texttt{Infer}#1}}

\newcommand{\oO}[1]{\ensuremath{\bm{O}_{#1}}}
\newcommand{\pP}[1]{\ensuremath{\bm{P}_{#1}}}

\newcommand{\tyR}[1]{\ensuremath{\tyname{R}_{#1}}}

\newcommand{\tytuple}[1]{\ensuremath{\langle#1\rangle}}
\newcommand{\tyunit}{\ensuremath{\langle\rangle}}
\newcommand{\tydist}[1]{\ensuremath{\tyname{Dist}(#1)}}

\newcommand{\moda}{\ensuremath{\textsc{A}}}
\newcommand{\modb}{\ensuremath{\textsc{P}}}
\newcommand{\modc}{\ensuremath{\textsc{N}}}
\newcommand{\modd}{\ensuremath{\textsc{Det}}}
\newcommand{\modr}{\ensuremath{\textsc{Rnd}}}

\newcommand{\tm}[1]{\ensuremath{t_{#1}}}
\newcommand{\ty}[1]{\ensuremath{T_{#1}}}
\newcommand{\mc}[1]{\ensuremath{c_{#1}}}
\newcommand{\mcv}[1]{\ensuremath{\bar{c}_{#1}}}
\newcommand{\me}[1]{\ensuremath{e_{#1}}}
\newcommand{\mE}{\ensuremath{E}}
\newcommand{\md}{\ensuremath{d}}
\newcommand{\mr}[1]{\ensuremath{r_{#1}}}
\newcommand{\ms}[1]{\ensuremath{s_{#1}}}
\newcommand{\mv}[1]{\ensuremath{v_{#1}}}
\newcommand{\mw}[1]{\ensuremath{w_{#1}}}
\newcommand{\mf}[1]{\ensuremath{f_{#1}}}
\newcommand{\mpp}{\ensuremath{p}}
\newcommand{\mDist}{\ensuremath{\texttt{D}}}

\newcommand{\eabs}[2]{\ensuremath{\lambda #1.\; #2}}
\newcommand{\etuple}[1]{\ensuremath{\langle#1\rangle}}
\newcommand{\eunit}{\ensuremath{\langle\rangle}}
\newcommand{\eproj}[3]{\ensuremath{\pi^{#1}_{#2}\; #3}}

\newcommand{\kif}{\kw{if}}
\newcommand{\kthen}{\kw{then}}
\newcommand{\kelse}{\kw{else}}
\newcommand{\eif}[3]{\ensuremath{\kif{}\;#1 > 0\;\kthen{}\;#2\;\kelse{}\;#3}}
\newcommand{\kassume}{\kw{assume}}
\newcommand{\eassume}[1]{\ensuremath{\kassume{}\; #1}}
\newcommand{\kweight}{\kw{weight}}
\newcommand{\eweight}[1]{\ensuremath{\kweight{}\; #1}}

\newcommand{\kinfer}{\kw{infer}}
\newcommand{\einfer}[1]{\ensuremath{\kinfer{}\; #1}}
\newcommand{\kdiff}[1]{\ensuremath{\kw{diff}_{#1}}}
\newcommand{\kdiffone}[1]{\ensuremath{\kw{diff1}_{#1}}}
\newcommand{\ediff}[3]{\ensuremath{\kdiff{#1}\; #2\; #3}}
\newcommand{\ediffone}[3]{\ensuremath{\kdiffone{#1}\; #2\; #3}}
\newcommand{\ksolve}{\kw{solve}}
\newcommand{\esolve}[3]{\ensuremath{\ksolve\; #1\; #2\; #3}}
\newcommand{\klet}{\kw{let}}
\newcommand{\kin}{\kw{in}}
\newcommand{\elet}[2]{\ensuremath{\klet\; #1=#2}}
\newcommand{\eletin}[2]{\ensuremath{\elet{#1}{#2}\; \kin}}

\newcommand{\kobserve}{\kw{observe}}
\newcommand{\eobserve}[2]{\ensuremath{\kobserve\; #1\; #2}}

\newcommand{\idname}[1]{\ensuremath{\textsc{#1}}}

\newcommand{\ipdf}[1]{\ensuremath{\text{pdf}_{#1}}}
\newcommand{\iwiener}[1]{\ensuremath{\text{wiener}_{#1}}}

\newcommand{\hole}[1]{\ensuremath{[#1]}}

\newcommand{\den}[1]{\ensuremath{\llbracket #1 \rrbracket}}

\newcommand{\ldotsTwo}{%
  \mathinner{{\ldotp}{\ldotp}}%
}
\newcommand{\dotdot}[2]{%
  #1\ldotsTwo#2%
}
\makeatletter
\def\moverlay{\mathpalette\mov@rlay}
\def\mov@rlay#1#2{\leavevmode\vtop{%
   \baselineskip\z@skip \lineskiplimit-\maxdimen
   \ialign{\hfil$\m@th#1##$\hfil\cr#2\crcr}}}
\newcommand{\charfusion}[3][\mathord]{
    #1{\ifx#1\mathop\vphantom{#2}\fi
        \mathpalette\mov@rlay{#2\cr#3}
      }
    \ifx#1\mathop\expandafter\displaylimits\fi}
\makeatother

\newcommand{\stan}{\texttt{Stan}}
\newcommand{\zelus}{\texttt{Zelus}}
\newcommand{\probzelus}{\texttt{ProbZelus}}

\newcommand{\pymc}{\texttt{PyMC}}

\newcommand{\julia}{\texttt{Julia}}
\newcommand{\theano}{\texttt{Theano}}

\usepackage{listings}
\usepackage{xcolor}

\def \dpplfontsize {\small}

\def \mcorefont {\fontfamily{pcr}\selectfont\dpplfontsize}

\definecolor{codegreen}{rgb}{0,0.6,0}
\definecolor{codegray}{rgb}{0.5,0.5,0.5}
\definecolor{codepurple}{rgb}{0.58,0,0.82}

\lstdefinelanguage{DPPL}{
  morekeywords={Lam,con,else,end,fix,if,in,lam,lang,let,match,recursive,sem,syn,then,type,use,utest,with,assume,solveode,observe,infer},
  otherkeywords={->},
  keywordstyle=\color{blue}\bfseries,
  morekeywords=[2]{Float, FloatP, FloatN, Int},
  keywordstyle=[2]\color{codegreen}\bfseries,
  morecomment=[l][\color{codegray}]{--},
  morestring=[b]",
  stringstyle=\color{codepurple},
  sensitive=true,
  basicstyle=\mcorefont,
  breaklines=true,
  escapeinside={(*@}{@*)},
  numbers=left,
  stepnumber=1,
  numberstyle=\tiny,
  xleftmargin=2em, % numbers will be in the margins
  columns=fixed, % same width for all characters
  showstringspaces=false,
  mathescape=true,
  breaklines=true,
  breakatwhitespace=true,
  mathescape=true,
  showstringspaces=false
}

\newcommand{\dpplinline}[1]{\lstinline[language=DPPL]|#1|}

\lstdefinelanguage{PT}{
  morekeywords={periodic,read,write,to,offset,update,infer},
  otherkeywords={->},
  keywordstyle=\color{ckeywords}\bfseries,
  morekeywords=[2]{mexpr,include,never},
  keywordstyle=[2]\color{cwarnings},
  morecomment=[l][\color{ccomments}]{--},
  morestring=[b]",
  stringstyle=\color{cstrings},
  sensitive=true,
  basicstyle=\mcorefont,
  breaklines=true,
  escapeinside={(*@}{@*)},
  numbers=left,
  stepnumber=1,
  numberstyle=\tiny,
  xleftmargin=2em, % numbers will be in the margins
  columns=fixed, % same width for all characters
  showstringspaces=false,
  mathescape=true,
  breaklines=true,
  breakatwhitespace=true,
  mathescape=true,
  showstringspaces=false
}

\begin{document}

%%
%% The "title" command has an optional parameter,
%% allowing the author to define a "short title" to be used in page headers.
\title{\Cdppl{}: Towards a Sound Composition of Differentiation, ODE Solving, and Probabilistic Programming}

%%
%% The "author" command and its associated commands are used to define
%% the authors and their affiliations.
%% Of note is the shared affiliation of the first two authors, and the
%% "authornote" and "authornotemark" commands
%% used to denote shared contribution to the research.
\author{Oscar Eriksson}
\authornote{Both authors contributed equally to this research.}
\email{oerikss@kth.se}
\orcid{0009-0001-0678-549X}
\affiliation{%
	\institution{EECS and Digital Futures, KTH Royal Institute of Technology}
	\city{Stockholm}
	\country{Sweden}
}

\author{Anders Ågren Thuné}
\authornotemark[1]
\email{anders.agren-thune@it.uu.se}
\orcid{0009-0008-0847-5373}
\affiliation{%
	\institution{Department of Information Technology, Uppsala University}
	\city{Stockholm}
	\country{Sweden}
}

\author{Johannes Borgström}
\email{johannes.borgstrom@it.uu.se}
\orcid{0000-0001-5990-5742}
\affiliation{%
	\institution{Department of Information Technology, Uppsala University}
	\city{Uppsala}
	\country{Sweden}
}

\author{David Broman}
\orcid{0000-0001-8457-4105}
\email{dbro@kth.se}
\affiliation{%
	\institution{EECS and Digital Futures, KTH Royal Institute of Technology}
	\city{Stockholm}
	\country{Sweden},
}

%%
%% By default, the full list of authors will be used in the page
%% headers. Often, this list is too long, and will overlap
%% other information printed in the page headers. This command allows
%% the author to define a more concise list
%% of authors' names for this purpose.
% \renewcommand{\shortauthors}{Trovato and Tobin, et al.}

%%
%% The abstract is a short summary of the work to be presented in the
%% article.
\begin{abstract}
	In recent years, there has been extensive research on how to extend
	general-purpose programming language semantics with domain-specific
	modeling constructs. Two areas of particular interest are (i)
	universal probabilistic programming where Bayesian probabilistic
	models are encoded as programs, and (ii) differentiable programming
	where differentiation operators are first class or differential
	equations are part of the language semantics. These kinds of languages
	and their language constructs are usually studied separately or
	composed in restrictive ways. In this paper, we study and formalize the
	combination of probabilistic programming constructs, first-class
	differentiation, and ordinary differential equations in a higher-order setting. We
	propose formal semantics for a core of such differentiable probabilistic programming language (DPPL), where the type system
	tracks random computations and rejects unsafe compositions during
	type checking. The semantics and its type system are formalized,
	mechanized, and proven sound in Agda with respect to abstract language constructs.

\end{abstract}

%%
%% Keywords. The author(s) should pick words that accurately describe
%% the work being presented. Separate the keywords with commas.

%%
%% This command processes the author and affiliation and title
%% information and builds the first part of the formatted document.
\maketitle

\section{Introduction}

Mathematical modeling languages and tools are vital in science and engineering, using methods such as Bayesian probabilistic modeling and inference~\cite{carpenter17,ghahramani2015probabilistic}, equation-based modeling and simulation~\cite{broman10,fritzson2014principles}, or differentiation as part of machine learning applications~\cite{baydin2018automatic,rumelhart1986learning}.
A recent trend is to augment standard programming languages with domain-specific language constructs, thus enabling highly expressive modeling within higher-order general-purpose languages.
For instance, \emph{probabilistic programming languages (PPLs)}~\citep{goodman08, murray18, wood14, goodman14, bingham19, cusumano-towner19, paquet21, lunden22, dash23, abril23} is a class of modeling languages where probabilistic models can be formulated as programs.
Another direction is incorporating differentiation primitives as first-class language constructs, resulting in \emph{differentiable programming} (cf. ~\citet{abadiplotkin19}), which has a long history that goes back to \emph{automatic differentiation} (AD) systems.
Differentiation in programming languages can take different forms;
in particular, machine learning and neural networks have reinvigorated research on the differentiation of programs (cf.~\citet*{baydin2018automatic}) as well as the inclusion of primitives for solving ordinary differential equations (ODEs) as part of the language (cf. ~\citet{benveniste18}).

Although the incorporation of probabilistic and differentiable primitives into general-purpose programming languages enables high modeling expressivity, a fundamental problem concerns the correctness of compositions:
\emph{how can program terms with both probabilistic and differentiation primitives be composed safely while still retaining sufficient expressivity?}
For instance, taking the derivative of a function that includes randomness is generally not well-defined, and solutions to ODEs have continuity requirements on their derivatives that are easily violated when random effects appear without any restriction in the model.
Issues for differentiability include that in programs with branching, the derivative in the analytic sense is sometimes only well-defined piecewise, and realizations of, e.g., the Wiener process are not even piecewise analytic.

Traditionally, the correctness of PPLs~\citep{borgstrom16,vakar19,lunden21,scibior18} and program differentiation~\cite{mazza21,abadiplotkin19,lee20,elliott18,krawiec22} has been studied separately.
For differentiation, the notion of a derivative is either extended to piecewise analytic functions or the language restricts primitive Boolean operations to where the derivative is well-defined.
There are a few works that study the combination of differentiation and probabilistic programming in a formal setting, showing the correctness of forward-mode AD of functions returning expectations~\citep{lew23}, and defining an extended notion of derivatives in the context of AD transformation of PPL programs~\citep{huot23}.
There is, however, no comprehensive treatment on the correctness of combining the three parts (i) higher-order universal PPLs with inference operators, (ii) first-class differentiation operators, and (iii) first-class constructs for solving ODEs.

This paper studies a \emph{differentiable probabilistic programming language (DPPL)} with (i) PPL constructs (\kinfer{}, \kassume{}, \kweight{}), (ii) first-class differentiation (a \kdiff{} construct), and (iii) ordinary differential equations (a \ksolve{} primitive) and how these can be composed in a higher-order language context.
Furthermore, we propose formal semantics for a higher-order DPPL calculus.
We take a different approach from previous work in that we develop a static semantics intended to separate deterministic and random parts of the program, as well as functions typed analytic, piecewise analytic, and non-differentiable.
A key idea of the static semantics is that the type system is based on effect \citep{lindley12, lucassen88, gaboardi16, leijen14} and coeffect systems \citep{gaboardi16, petricek14, brunel14}.
The dynamic semantics builds on previous work on PPL \citep{borgstrom16} and AD \citep{lee20} correctness, as well as ODE theory \citep{kelley10}.

More specifically, the contributions are as follows:
\begin{itemize}
	\item We illustrate key ideas of composition through a motivating example (\Cref{sec:motivation1}).
	\item We develop formal semantics with a novel type system for a core DPPL, where the type system guarantees that deterministically typed terms do not express random computations, that analytic derivatives are only taken on functions of analytic type, and that intensional derivatives are only taken on functions of piecewise analytic type (Sections \ref{sec:syntax} and \ref{sec:formalization}).
	\item We formalize and mechanize proofs in Agda, proving type safety under the assumption of correct implementations of primitive operations and that deterministically typed terms cannot affect the random state (\Cref{sec:mechanization}).
	\item We implement the core DPPL for an extended subset of an existing compiled universal PPL (\Cref{sec:implementation}), and evaluate the implementation on several case studies (\Cref{sec:case-studies}).
\end{itemize}

\section{Motivation}\label{sec:motivation1}

To motivate the relevance of \cdppl{}, this section first gives a general overview of the case studies of \Cref{sec:case-studies}; and then presents a part of one of the case studies in more detail.

\begin{figure*}[t]
	\includegraphics[width=0.6\textwidth]{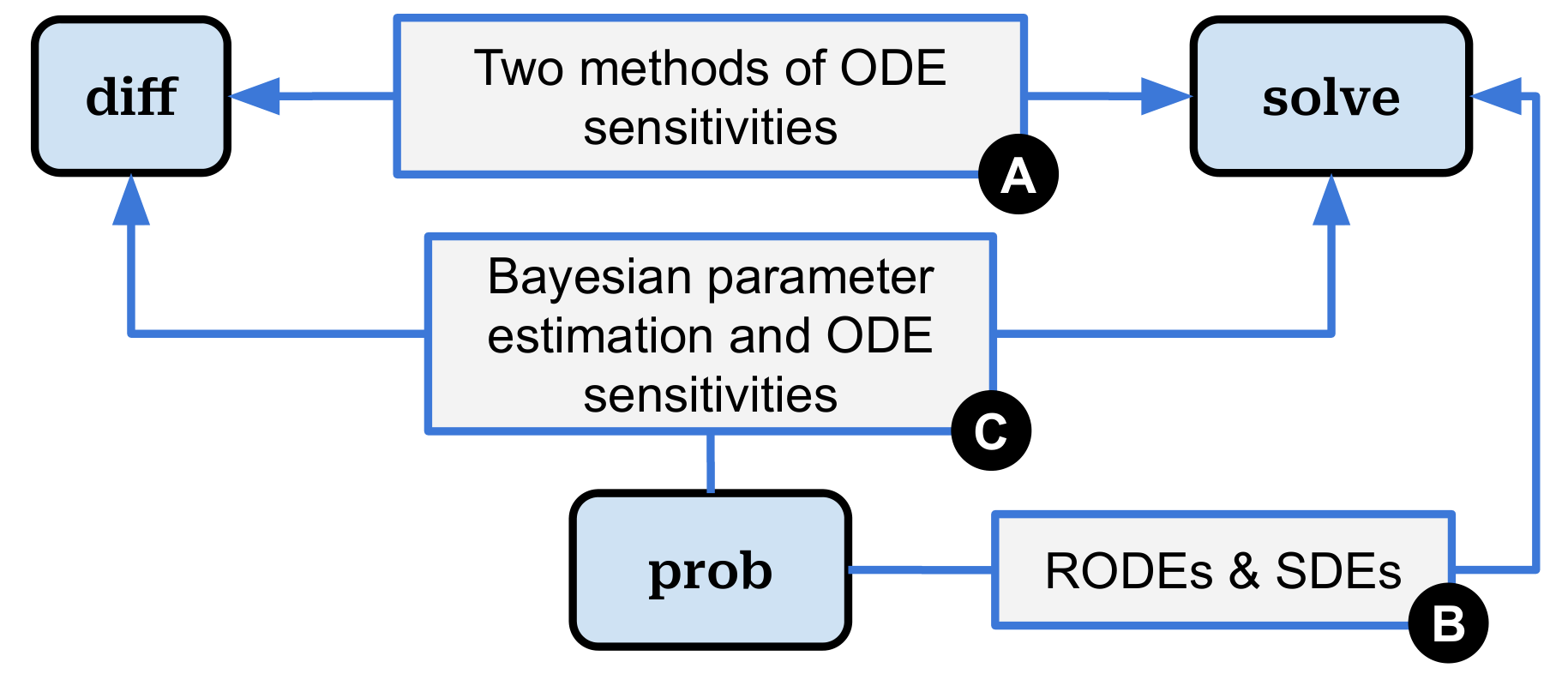}
	\caption{Overview of examples illustrating the compositions of \kdiff{}, \ksolve{}, and the probabilistic constructs (\textbf{prob}): \kinfer{}, \kassume{}, and \kweight{}. The direction of the arrows gives a notion of nesting where the arrow points to the nested (inner) construct.}
	\label{fig:overview}
\end{figure*}

\Cref{fig:overview} gives an overview of three examples (A to C) that illustrate the relevance of the composition of differential and probabilistic constructs.
Example A (see \Cref{sec:motivation:sens} in the case study for details) illustrates two methods for computing the derivative of an ODE solution w.r.t. to its parameters: by differentiating \ksolve{} itself or by augmenting the original ODE with sensitivity equations by differentiating the ODE model rather than its solution.
These derivatives are called the sensitivities of the ODE and are an important property for analyzing an ODE model, and they appear in the gradient computation for objective functions that involve ODE solutions.
Example B (details in the case study
in \Cref{sec:motivation:rode}) shows how we can compose \ksolve{} and probabilistic constructs to formulate so-called Random ODEs, which have several important applications \citep{imkeller01, neckel17}.
Finally, we introduce example C, which composes all three kinds of constructs, \kdiff{}, \ksolve{}, and the probabilistic constructs \kassume{}, \kobserve{}, and \kinfer{} in the next paragraph.

\begin{figure*}[b]
	\begin{subfigure}[b]{0.57\textwidth}
		\begin{lstlisting}[language=DPPL, basicstyle=\fontfamily{pcr}\selectfont\footnotesize, xleftmargin=0em]
let model = lam t : (). $\label{line:big:one}$
  let $\theta$ = assume (Gaussian 1. 1.) in $\label{line:big:two}$
  let $\sigma$ = assume (Beta 2. 2.) in $\label{line:big:three}$
  let ys = solve (ode $\theta$) y0 nData hData in $\label{line:big:four}$
  iter (lam t : ((FloatN, FloatN), FloatN). $\label{line:big:five}$
          match t with (y, o) in $\label{line:big:six}$
          observe o (Gaussian (output y) $\sigma$)) $\label{line:big:seven}$
    (zip ys data); $\label{line:big:eight}$
  let f = lam x : FloatP. solve (ode x) y0 n h $\label{line:big:ten}$
  in diff1P f $\theta$ $\label{line:big:eleven}$
in let dist = infer ($\ldots$) model $\label{line:big:twelve}$
    \end{lstlisting}
	\end{subfigure}%
	\begin{subfigure}[b]{0.35\textwidth}
		\includegraphics[width=\textwidth]{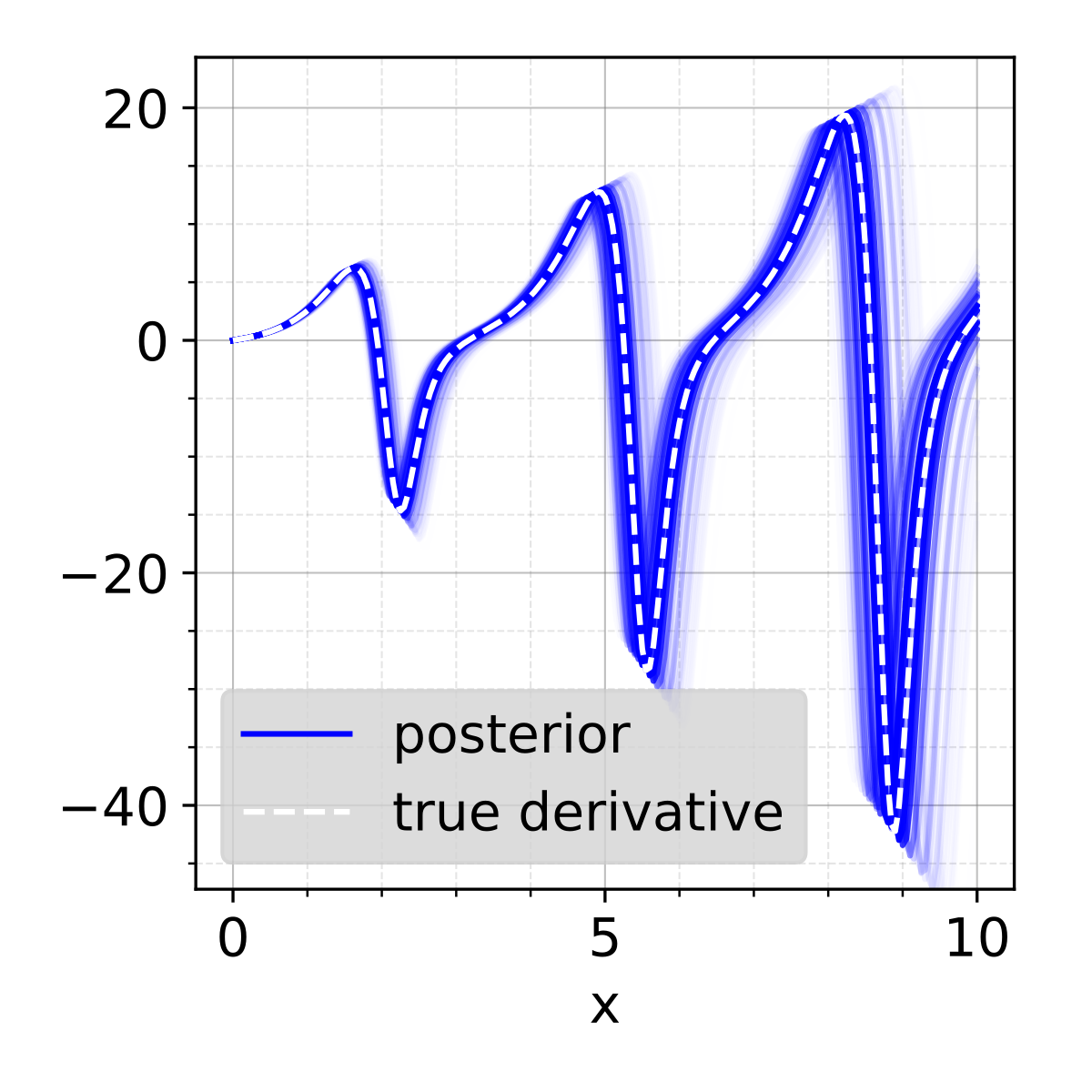}
	\end{subfigure}
	\caption{Left: a probabilistic model in our \cppl{} language, based on \miking{} \cppl{}, that includes parameter estimation from noisy data for a dynamical system as well as ODE sensitivity computation by program differentiation. Right: a graph of the distribution of traces for the sensitivity of the density of prey (blue) and the trace of the sensitivity for the true value of $\theta=1.5$ (dashed white) over time ($x$). The shade of blue is proportional to the probability of a particular trace (darker means more probable).}
	\label{fig:big-example}
\end{figure*}

\Cref{fig:big-example} shows a code snippet from our implementation of example C,
which expresses the posterior distribution of the sensitivities of the solutions to the Lotka-Volterra predator-prey ODE model (\Cref{apdx:lotak-volterra}) w.r.t. the unknown prey per-capita growth rate parameter ($\theta$).
This ODE models prey and predator densities over time.
The posterior is conditioned on noisy observations from the true system.
The model is implemented as a computation in a thunk and the argument on line \ref{line:big:one} is unused.
Lines \ref{line:big:two} and \ref{line:big:three} encode our prior belief for the unknown parameter $\theta$ and measurement noise (\(\sigma\)) using the \kassume{} construct.
The line \ref{line:big:four} produces a solution trace from the parametrized ODE model (named \dpplinline{ode}, which is not shown in the figure for space reasons, see \Cref{apdx:lotak-volterra} for its definition) applied to \(\theta\) with known initial values (\dpplinline{y0}).
The variable \dpplinline{nData} is the number of measurements from the true system, and the variable \dpplinline{hData} is the time interval between two successive measurements.
The function \dpplinline{solve} internally uses an implementation of the \ksolve{} construct.
Lines \ref{line:big:five} to \ref{line:big:eight} condition the model on the observed measurements (\dpplinline{data}) with the \kobserve{} construct (line \ref{line:big:seven}).
The output function (\dpplinline{output}) models what we can observe in the true system; in this case, we can observe the prey density but not the predator density.
In a language that includes both \ksolve{} and \kdiff{}, it is natural to compute ODE sensitivities by differentiating the ODE solution w.r.t. its parameters.
Lines \ref{line:big:ten} to \ref{line:big:eleven} produce the sensitivities by differentiating (\dpplinline{diff1P}, which is essentially \kdiff{} for functions with a scalar parameter with additional type annotations)
the ODE solution w.r.t. \(\theta\) (the variables \dpplinline{n} and \dpplinline{h} define the number of time points and the time between time points for the ODE solution trace, respectively).
Finally, line \ref{line:big:twelve} infers a posterior distribution (\dpplinline{dist}) of sensitivity traces plotted on the right-hand side in \Cref{fig:big-example}, where \dpplinline{($\dots$)} elides inference algorithm-specific parameters.

As a final note on types in \Cref{fig:big-example}. Our type system decorates types representing real numbers (\dpplinline{Float} in the implementation) with coeffect modifiers: \modb{}, which stands for \emph{piece-wise analytic under analytic partitioning} (PAP) \citep{lee20}, and \modc{}, which stands for non-differentiable. The third and final modifier is \moda{}, which stands for analytic and is not part of this example. Parameters on line \ref{line:big:five} are annotated with \dpplinline{FloatN}, which types this function as not having a well-defined derivative because \dpplinline{t} is involved in the conditioning on data on line \ref{line:big:seven}.
On the other hand, the function \dpplinline{f} on line \ref{line:big:ten} is differentiable, and its parameter is typed \dpplinline{FloatP} and, consequently, \dpplinline{diff1P f} is well-typed.
The type system tracks and propagates these modifiers in a simplified coeffect system that ensures consistent type annotations.

\section{Differentiable Probabilistic Programming}\label{sec:syntax}

\begin{figure*}[t]
	\begin{tabular}{c}
		$\mc{} \in\{\moda{},\modb{},\modc{}\}\quad\md \in\{\moda{},\modb{}\}\quad\me{} \in\{\modd{},\modr{}\}\quad x\in\dIdent{}\quad\mr{} \in \dR{}\quad\mpp{} \in [0,1]\quad i,n,m \in \dN{}$ \\[0.2cm]
		\begin{tabular}{lcl}
			\ty{}       & $\Coloneqq$ & $\tyR{\mc{}} ~\mid~ \ty{1}\to^{\me{}}\ty{2} ~\mid~ \tytuple{\ty{1}, \dots, \ty{n}} ~\mid~ \tydist{\ty{}}$ \hfill \textit{Types}                                                                       \\
			\tm{},\mf{} & $\Coloneqq$ & $x ~\mid~ \lambda x:\ty{}.\;\tm{} ~\mid~ \mf{}\; \tm{} ~\mid~ \phi(\tm{1},\ldots,\tm{|\phi|}) ~\mid~ \mr{} ~\mid~ \etuple{\tm{1},\ldots,\tm{n}} ~\mid~ \eproj{n}{i}{\tm{}}\quad$ \hfill\textit{Terms} \\
			{}          & ~$\mid$~    & $\eif{\tm{}}{\tm{1}}{\tm{2}} ~\mid~ \mDist(\tm{1},\ldots,\tm{|\mDist|}) ~\mid~ \kassume{}\; \tm{} ~\mid~ \kweight\; \tm{}$                                                                            \\
			{}          & ~$\mid$~    & $\kinfer\; \mf{} ~\mid~ \kdiff{\md}\; \mf{}\; \tm{} ~\mid~ \ksolve{}\; \mf{}\; \tm{1}\; \tm{2}$                                                                                                       \\[0.2cm]
			\mDist      & $\Coloneqq$ & $\mathcal{N} \mid \mathcal{B}  \mid \mathcal{W} \mid \ldots$ \hfill \textit{Primitive distributions}                                                                                                  \\
			$\phi$      & $\Coloneqq$ & $+ \mid - \mid \cdot \mid \div \mid \sin \mid \cos \mid \ipdf{\mDist{}'} \mid \iwiener{\mpp{}} \mid \ldots$ \hfill \textit{Primitive functions}
		\end{tabular}
	\end{tabular}
	\caption{The syntax of \cdppl{}.}
	\label{fig:syntax}
\end{figure*}

This section introduces \cdppl{}.
\Cref{fig:syntax} summarizes its syntax, and we discuss the types in \Cref{sec:syntax-types}, the basic language syntax in \Cref{sec:syntax-basic}, probabilistic terms in \Cref{sec:syntax-ppl-terms}, and differential terms in \Cref{sec:syntax-diff-terms}.

\subsection{Types, random effect modifiers, and differential coeffect modifiers}
\label{sec:syntax-types}

\Cref{fig:syntax} defines the syntax of \cppl{}, where \dIdent{} is the set of variable names. The main goal of the type system is to track randomness, using an effect modifier \me{} on arrow types \(\ty{1}\to^{\me{}}\ty{2}\), and differentiability, using a co-effect modifier \mc{} on the real number type \tyR{\mc{}} to ensure that the probabilistic and differential primitives only receive compatible arguments.

The effect modifier \me{} can be either deterministic (\modd{}) or random (\modr{}), where a term typed \(\ty{1}\to^{\modr{}}\ty{2}\) expresses a random function, where the return value may depend on a finite (but a priori unbounded) number of random draws from the primitive distributions discussed below.
The modifier \mc{} can be either analytic (\(\mc{}=\moda{}\)),  PAP (\(\mc{} = \modb{}\)), or non-differentiable (\(\mc{}=\modc{}\)).
The types \tyR{\moda{}} and \tyR{\modb{}} correspond to the smooth and non-smooth types \dR{} and \(\dR{}^{*}\), respectively, in \citet{lew23}.

Here, a term typed \(\tyR{\moda{}}\to^{\modd{}}\tyR{\mc{}}\) expresses a scalar function typed as analytic, which naturally is deterministic.
Similarly, terms typed \(\tyR{\modb{}}\to^{\modd{}}\tyR{\mc{}}\) and \(\tyR{\modc{}}\to^{\modd{}}\tyR{\mc{}}\) express deterministic functions typed PAP and non-differentiable, respectively.

In the static semantics, \Cref{sec:static-semantics}, \me{} is tracked in a lightweight effect system \citep{lucassen88, lindley12, leijen14}, while \mc{} is tracked in a simplified coeffect system \citep{petricek14, brunel14, gaboardi16}.
A key difference to standard coeffect systems is that we attach these differentiability modifiers to the base type \tyR{} rather than to arrow types and variables in the type environment.
The reason for this is two-fold: (i) differentiability is not applicable to all types, and (ii) it allows for less conservative reasoning on differentiability for functions with parameters of tuple type.

The remaining types are tuple types, where \tyunit{} is the empty tuple;
and distribution types, where \tydist{\ty{}} means a probability distribution whose support has type \ty{}.
For convenience we write a homogeneous \(n\)-tuple \(\tytuple{\ty{}, \dots, \ty{}}\) as \(\ty{}^n\).
The syntactic sugar \(\tyR{\mcv{}}^n\) denotes an \(n\)-tuple \(\tytuple{\tyR{\mc{1}}, \dots, \tyR{\mc{n}}}\) of real numbers with $n$ possibly different modifiers \((\mc{1},\ldots,\mc{n}) \eqcolon \mcv{}\).
To reduce the notational burden, we make no distinction between \tytuple{\ty{}} and \ty{}.

\subsection{The Extended \(\lambda\)-calculus, Primitive Functions and Distributions, and Syntactic Sugar}
\label{sec:syntax-basic}

For terms \tm{} and \mf{}, we reserve \mf{} for terms of function type. We will sometimes overload \mf{} to functions in the mathematical sense.
Terms consist of the simply typed \(\lambda\)-calculus extended with the standard primitive function application, where \(\phi\) ranges over primitive functions and \(|\phi|\) is its arity; real numbers; tuples; tuple projection, where we sometimes omit the tuple size \(n\); and if-then-else-terms.
We omit boolean terms in our calculus for simplicity, but they are present in our implementation.

In our examples we use standard syntactic sugar for let expressions and sequencing, where we informally omit the type annotation.
We write \(\mr{}^n\) for an n-tuple of real numbers and also use sugar for pattern matching on tuples in both function parameters and let expressions.

Primitive functions (\(\phi\)) include elementary functions and probability density functions for the primitive distributions that possess them (\(\mDist{}' = \{\mathcal{N}, \mathcal{B}\}\)).
The primitive functions also include a family \iwiener{\mpp{}} of realizations of the Wiener process (discussed in \cref{sec:syntax-diff-terms}), where \(\mpp{}\) is an index into the set of all such realizations. More primitive functions can be added. We write primitive function applications with infix notation when appropriate.
Moreover, we reduce the notational burden and make no distinction between \(\eproj{1}{1}\; \tm{}\) and \tm{}, and \etuple{\tm{}} and \tm{}.

The remaining language primitives either relate to the probabilistic or the differential aspect of \cdppl{}.

\subsection{Probabilistic Terms}
\label{sec:syntax-ppl-terms}

There are four primitives in the probabilistic aspect, beginning with primitive distributions \mDist{} applied to a tuple of their parameters.
Primitive probability distributions \mDist{} includes: \(\mathcal{N}\), the normal distribution; \(\mathcal{B}\), the beta distribution; and \(\mathcal{W}\), the Wiener process. Other distributions can be added.

We model a prior distribution using \eassume{\tm{}}, which draws a sample of type \ty{} from a distribution \(\tm{}:\tydist{\ty{}}\) while evidence is modeled with
\eweight{\mr{}}: a side-effecting operation that multiplies the weight of the current runtime state by \mr{} (see Definition \ref{def:sampling-semantics}).
Given a probabilistic model \(\mf{}:\tyunit \to^{\modr{}}\ty{}\), the term \einfer{\mf{}} infers a posterior distribution of type \tydist{\ty{}}.

The syntactic sugar \(\eobserve{\tm{}}{\mDist'(\tm{1},\ldots,\tm{|\mDist'|})} \coloneq \kweight\; \ipdf{\mDist'}(\tm{1},\ldots,\tm{|\mDist'|},\tm{})\) conditions a probabilistic model on an observation \tm{} distributed according to the given distribution, where \ipdf{\mDist'} is the probability density function of \(\mDist'\).

As an example of a probabilistic term, consider the scalar regression model \(y = f(x, \theta) + \nu\), where \(y \) are observable outputs, \(x\) are known inputs, \(\theta\) is an unknown parameter independent of \(x\), and \(\nu \sim \mathcal{N}(0,1)\) is Gaussian measurement noise.
Moreover, assume
we have access to data in the form of an \(n\)-tuple, \(\idname{Data}\), of input-output pairs.
Then, omitting type annotation for brevity; using syntactic sugar for let bindings, tuple iteration and destruction, and sequencing; and the term
\begin{equation}
	\label{eq:basic-regression-example}
	\begin{aligned}
		 & \klet\; \idname{model} = \lambda \_ .                                                                                                                                                                                           \\
		 & \quad \eletin{\theta}{\eassume{\mathcal{N}(1,1)}}                                                                                                                                                                               \\
		 & \quad \eobserve{(\eproj{n}{1}\; \idname{Data})}{\mathcal{N}(\mf{}\; \etuple{\eproj{n}{1}\; x, \theta}, 1)};\ldots\eobserve{(\eproj{n}{n}\; \idname{Data})}{\mathcal{N}(\mf{}\; \etuple{\eproj{n}{1}\; x, \theta}, 1)};\; \theta \\
		 & \kin\; \kinfer\; \idname{model}
	\end{aligned}
\end{equation}
express the posterior distribution of the parameter \(\theta\) in a Bayesian regression model, where we model a prior belief that \(\theta\) is distributed according to the standard normal distribution and condition the posterior distribution on data by observing the input-output pairs of \(\idname{Data}\).
The infix operator (\(;\)) denotes expression sequencing, and the model returns the random variable \(\theta\) and \kinfer{} produces a posterior distribution of \(\theta\).
In \eqref{eq:basic-regression-example}, \(\idname{Model}\) is typed \(\tyunit{} \to^{\modr{}}\tyR{\modc{}}\), denoting a random function because its body contains the PPL constructs \kassume{} and \kweight{}, and \eqref{eq:basic-regression-example} is typed \(\tydist{\tyR{\modc{}}}\).

\subsection{Differential Terms}
\label{sec:syntax-diff-terms}

For simplicity we here discuss the scalar derivative \ediffone{\md{}}{\mf{}}{\mr{}}, which expresses the derivative of a deterministic function \(\mf{} : \tyR{\md{}}\to^{\modd{}}\tyR{\mc{}}\) at the point \mr{}.
Note that the modifier \md{} (\Cref{fig:syntax}) is on the function argument and not the result type: it denotes that the term \kdiffone{\md{}} expresses the derivative of either an analytically typed function (\md{}=\moda{}), or the intensional derivative of an PAP typed function (\md{}=\modb{}).
The general differentiation term \ediff{\md{}}{\mf{}}{\mr{}^n}, which expresses the total derivative.

Taking the analytic derivative of a quadratic polynomial,  the term
\begin{equation}
	\label{eq:motivation:a-derivative}
	\eletin{y}{\eabs{x:\tyR{\moda{}}}{x \cdot x + x}}\; \ediffone{\moda{}}{y}{\mr{}}
\end{equation}
is well-typed because \(y\) is a composition of primitive analytic functions (scalar addition (\(+\)) and multiplication (\(\cdot\))).
In contrast, since the absolute value function does not have an analytic derivative at zero, the following term is not typeable due to the \moda{} annotation on the argument:
\begin{equation}
	\label{eq:motivation:b-derivative}
	\eabs{x:\tyR{\moda{}}}{\eif{x}{x}{0-x}}.
\end{equation}
However, \eqref{eq:motivation:b-derivative} has a well-defined intensional derivative since it has a PAP representation (and is, therefore, a PAP function) \citep{lee20}.
We can show this by dividing the domain of the function into two analytic intervals, \(\mathbb{R}_{\leq0}\) and \(\mathbb{R}_{>0}\), associated with the analytic functions \(x \mapsto -x\) and \(x \mapsto x\), respectively.
The notion of an intensional derivative is important because it captures the semantics of almost all derivatives expressed using AD~\citep{huot23},
so
\begin{equation}\label{eq:if-example}
	\eletin{y}{\eabs{x:\tyR{\modb{}}}{\eif{x}{x}{0-x}}}\; \ediffone{\modb{}}{y}{\mr{}}
\end{equation}
is well-typed in our type system.
The type system also allows us to replace the modifier \moda{} in \eqref{eq:motivation:a-derivative} with modifier \modb{} since an analytic function is a PAP function with one analytic interval (the whole of \(\mathbb{R}\) in this case), and we also allow the use of an analytically typed function in contexts that expect a PAP typed function.

Our type system distinguishes between analytically typed and PAP typed functions by propagating coeffects and constraining the comparand \tm{} in
\(
\eif{\tm{}}{\tm{1}}{\tm{2}}
\)
to the type \tyR{\modb{}}.
Relating the probabilistic and the differential part of \cdppl{}, the term
\begin{equation}
	\eletin{y}{\eassume{\tm{}}}\; \ediffone{\md{}}{(\eabs{x:\tyR{\md{}}}{y})}{\mr{}}
\end{equation}
is well-typed when \tm{} has the appropriate type because once sampled, the sample \(y\) is deterministic in the body of the let term.

The primitive distributions (\Cref{fig:syntax}) include the distribution \(\mathcal{W}\) over Wiener processes, which allows us to express so-called Random ODEs (RODEs)~\citep{neckel17, imkeller01} by composing PPL constructs and \ksolve{} (see the example in \Cref{sec:motivation:rode}).
An interesting property of realizations of the Wiener process is that they are continuous deterministic scalar functions but nowhere differentiable, neither in the analytic nor the intensional sense.
We therefore type \(\mathcal{W}()\) as \(\tydist{\tyR{\modc{}}\to^{\modd{}}\tyR{\modc{}}}\), which gives its samples (or realizations) the type \(\tyR{\modc{}}\to^{\modd{}}\tyR{\modc{}}\). This type cannot be differentiated since the modifier on the function argument must be \moda{} or \modb{}.
On the other hand, as we discuss in \Cref{sec:static-semantics}, because we track the coeffect modifiers \mc{} on the type \tyR{}, the type system allows terms like
\begin{equation}
	\eletin{w}{\eassume{\mathcal{W}()}}\; \eletin{z}{\eabs{\etuple{x,y} : \tytuple{\tyR{\modc{}},\tyR{\moda{}}}}{w\; x + y}}\; \ediffone{\moda{}}{(\eabs{u:\tyR{\moda{}}}{z\; \etuple{\mr{}, u}})}{\mr{}},
\end{equation}
even though \(z\) includes a realization of the Wiener process, since \(z\) is typed analytic in its second parameter \(y\) (but not the first).

There is no limitation in using terms typed \tyR{\modc{}} in the comparand term of if-then-else expressions because the type system can safely coerce terms typed \tyR{\modc{}} to the required type \tyR{\modb{}}.

The term \esolve{\mf{}}{\mr{}^n}{\mr{}} expresses a solution \(y(x)\), at \(x = \mr{}\), to the Initial Value Problem (IVP) that consists of the ODE \(\frac{dy(x)}{dx} = f(x,y(x))\) (we consider vector-valued ODEs because they can express systems of ODEs and refer to \(x\) as time) and initial values \(y(0) = \mr{}^n\).
\Cdppl{} expects the term \mf{} to have the type \(\tytuple{\tyR{\mc{}},\tyR{\mcv{}}^n}\to^{\modd{}}\tyR{\mcv{}}^n\).
As an example, the term
\begin{equation}
	\esolve{(\eabs{\etuple{x,y}}{x - y})}{0}{\mr{}}
\end{equation}
express \(y(\mr{})\) for the solution \(y(x)\) to the IVP: \(\frac{dy(x)}{dx} = x - y(x); y(0) = 0\).
The \ksolve{} construct assumes an initial time \(0\).

Similar to \kdiff{}, the type system requires that the ODE right-hand side \mf{} is deterministic.
However, we do not restrict the modifiers \mc{} on the type of \mf{}.
The reason is that the existence of a unique solution is only guaranteed under certain additional assumptions, as we discuss in \Cref{sec:dynamic-semantics}.

As we illustrated in \Cref{sec:motivation1}, the composition of \kdiff{} and \ksolve{} allows us to express the sensitivities of IVP solutions (example A and C in \Cref{fig:overview}). Moreover, the composition of \ksolve{} and PPL constructs allows us to both express parameter estimation models of dynamical systems in a Bayesian setting (example C), as well as the previously mentioned RODEs (example B).

\section{Formalization}\label{sec:formalization}

This section formalizes the semantics of \cdppl{} introduced in \Cref{sec:syntax}.
In~\Cref{sec:static-semantics}, we define the static semantics; in~\Cref{sec:coeffect-compare} we compare our type-system to other effect systems;
and in~\Cref{sec:dynamic-semantics}, we define the dynamic semantics.

\subsection{Static Semantics}\label{sec:static-semantics}

\begin{figure*}[t]
	\begin{subfigure}{0.95\textwidth}
		\begin{tabular}{c}
			$\moda{}<\modb{}<\modc{} \qquad \modd{} < \modr{} \qquad \mc{1} \cdot \mc{2} \coloneq \max{(\mc{1},\mc{2})}$                                                                                                                                                                                                                               \\[0.1cm]
			\begin{tabular}{rclrcl}
				$\mc{1} \leq \tyR{\mc{2}}$                   & $\coloneq$ & $\mc{1}\leq\mc{2}$                                           & $\mc{1} \cdot \tyR{\mc{2}}$                   & $\coloneq$ & $\tyR{\mc{1}\cdot\mc{2}}$                                \\
				$\mc{} \leq \tytuple{\ty{1}, \dots, \ty{n}}$ & $\coloneq$ & $\mc{} \leq \ty{1} \land \dots \land \mc{} \leq \ty{n}\quad$ & $\mc{} \cdot \tytuple{\ty{1}, \dots, \ty{n}}$ & $\coloneq$ & $\tytuple{\mc{} \cdot \ty{1},\dots, \mc{} \cdot \ty{n}}$ \\
				$\mc{} \leq \ty{}$                           & $\coloneq$ & $\top$ (for all other \ty{})                                 & $\mc{} \cdot \ty{}$                           & $\coloneq$ & \ty{} (for all other \ty{})
			\end{tabular} \\[0.5cm]
			$\Gamma \Coloneqq \{\} \mid \Gamma,x:\ty{} \qquad \mc{} \leq \{\} \coloneq \top \qquad \mc{} \leq (\Gamma,x:\ty{}) \coloneq (\mc{} \leq \Gamma) \wedge (\mc{} \leq \ty{})$
		\end{tabular}
		\label{fig:meta:coeffects-effects}
	\end{subfigure}
	\begin{subfigure}[b]{0.90\textwidth}
		\begin{gather*}
			\inferrule[]
			{\mc{}' \leq \mc{}}
			{\tyR{\mc{}} \leq: \tyR{\mc{}'}}\quad
			\inferrule[]
			{\ty{1}' \leq: \ty{1}\quad\ty{2} \leq: \ty{2}'\quad\me{} \leq \me{}'}
			{\ty{1}\to^{\me{}}\ty{2} \leq: \ty{1}'\to^{\me{}'}\ty{2}'}\quad
			\inferrule[]
			{\ty{i} \leq: \ty{i}' \quad i = \dotdot{1}{n}}
			{\tytuple{\ty{1}, \dots,\ty{n}} \leq: \tytuple{\ty{1}', \dots, \ty{n}'}}\quad
			\inferrule[]
			{\ty{} \leq: \ty{}'}
			{\tydist{\ty{}} \leq: \tydist{\ty{}'}}\quad
		\end{gather*}
	\end{subfigure}
	\caption{Type environment, operations over effects and coeffects, and the sub-typing relation.}
	\label{fig:meta}
\end{figure*}

Before we give type rules of \cdppl{}, we define the type environment, auxiliary operations, and relations over (co)effect modifiers, types, and type environments in \Cref{fig:meta}, where we consider the type environment \(\Gamma\) a set.
Multiplication and order over (co)effect modifiers reflect that larger modifiers dominate smaller modifiers when determining the context of differentiation or randomness.
E.g., for all \mc{}, \(\moda{} \cdot \mc{} = \mc{} \cdot \moda{} = \mc{}\) and \(\modc{} \cdot \mc{} = \mc{} \cdot \modc{} = \modc{}\).
Because we attach coeffect modifiers on types \tyR{}, we also extend coeffect ordering and multiplication to types and ordering on type environments.
The bottom definition for coeffect-type ordering and coeffect-type multiplication reflects that we only need to change and track the modifiers on types isomorphic to \(\tyR{}^n\).
We never allow differentiation of functions with parameters of arrow or distribution type and therefore do not need to track coeffects on these types in the type environment.

For the sub-typing relation, \Cref{fig:meta}, intuitively, the modifier \modc{} is less restrictive than modifiers \moda{} and \modb{}: we can use \modc{} in differential and non-differential contexts, but the opposite is false.
Similarly, the modifier \modd{} on arrow types is less restrictive than \modr{} in our semantics: we can use a deterministic function in places that expect a random function but not vice versa.
The reflexivity property \(\ty{} \leq: \ty{}\) follows from the definition in \Cref{fig:meta}.

\begin{figure*}[t]
	\begin{gather*}
		\inferrule[T-Sub]
		{\Gamma \vdash \tm{} :^{\me{}} \ty{} \quad \ty{} \leq: \ty{}' \quad \me{} \leq \me{}'}
		{\Gamma \vdash \tm{} :^{\me{}'} \ty{}'}\quad
		\inferrule[T-Promote]
		{\Gamma \vdash \tm{} :^{\me{}} \ty{} \quad \mc{} \leq \Gamma}
		{\Gamma \vdash \tm{} :^{\me{}} \mc{}\cdot\ty{}}\quad
		\inferrule[T-Weaken]
		{\Gamma \vdash \tm{} :^{\me{}} \ty{} \quad
		\Gamma \subseteq \Gamma'}
		{\Gamma' \vdash \tm{} :^{\me{}} \ty{}}\\[2mm]
		\inferrule[T-Var]
		{}
		{x:\ty{} \vdash x :^{\modd{}} \ty{}}\quad
		\inferrule[T-Abs]
		{\Gamma,x:\ty{1} \vdash \tm{} :^{\me{}} \ty{2}}
		{\Gamma \vdash \lambda x:\ty{1}.\; \tm{} :^{\modd{}} \ty{1} \to^{\me{}} \ty{2}}\quad
		\inferrule[T-App]
		{\Gamma \vdash \mf{} :^{\me{}} \ty{1} \to^{\me{}} \ty{2} \quad \Gamma \vdash \tm{} :^{\me{}} \ty{1}}
		{\Gamma \vdash \mf{}\; \tm{} :^{\me{}} \ty{2}}\\[2mm]
		\inferrule[T-PrimApp]
		{\intrinsicType(\phi) = (\ty{1}, \ldots, \ty{|\phi|}, \ty{}) \quad \Gamma \vdash \tm{i} :^{\me{}} \ty{i} \quad i \in \dotdot{1}{|\phi|}}
		{\Gamma \vdash \phi(\tm{1}, \ldots, \tm{|\phi|}) :^{\me{}} \ty{}}\quad
		\inferrule[T-Real]
		{}
		{\Gamma \vdash \mr{} :^{\modd{}} \tyR{\modc{}}}\\[2mm]
		\inferrule[T-Tuple]
		{\Gamma \vdash \tm{i} :^{\me{}} \ty{i} \quad i \in \dotdot{1}{n}}
		{\Gamma \vdash \etuple{\tm{1}, \ldots, \tm{n}} :^{\me{}} \tytuple{\ty{1}, \dots, \ty{n}}}\quad
		\inferrule[T-Proj]
		{\Gamma \vdash \tm{} :^{\me{}} \tytuple{\ty{1},\dots,\ty{n}} \quad i \in \dotdot{1}{n}}
		{\Gamma \vdash \eproj{n}{i}{\tm{}} :^{\me{}} \ty{i}}\\[1mm]
		\inferrule[T-If]
		{\Gamma \vdash \tm{} :^{\me{}} \tyR{\modb{}} \quad \Gamma \vdash \tm{1} :^{\me{}} \ty{} \quad \Gamma \vdash \tm{2} :^{\me{}} \ty{}}
		{\Gamma \vdash \eif{\tm{}}{\tm{1}}{\tm{2}} :^{\me{}} \ty{}}\\[2mm]
		\inferrule[T-PrimDist]
		{\distType(\mDist{}) = (\ty{1}, \ldots, \ty{|\mDist{}|}, \ty{}) \quad \Gamma \vdash \tm{i} :^{\me{}} \ty{i} \quad i \in \dotdot{1}{|\mDist{}|}}
		{\Gamma \vdash \mDist{}(\tm{1}, \ldots, \tm{|\mDist{}|}) :^{\me{}} \tydist{\ty{}}}\\[2mm]
		\inferrule[T-Assume]
		{\Gamma \vdash \tm{} :^{\me{}} \tydist{\ty{}}}
		{\Gamma \vdash \kassume{}\; \tm{} :^{\modr{}} \ty{}}\quad
		\inferrule[T-Weight]
		{\Gamma \vdash \tm{} :^{\me{}} \tyR{\modc{}}}
		{\Gamma \vdash \kweight{}\; \tm{} :^{\modr{}} \tyunit}\quad
		\inferrule[T-Infer]
		{\Gamma \vdash \mf{} :^{\me{}} \tyunit \to^{\modr{}} \ty{}}
		{\Gamma \vdash \kinfer{}\; \mf{} :^{\me{}} \tydist{\ty{}}}\\[2mm]
		\inferrule[T-Diff]
		{\Gamma \vdash \mf{} :^{\me{}} \tyR{\md}^{n} \to^{\modd{}} \tyR{\mcv{}}^{m} \quad
		\Gamma \vdash \tm{} :^{\me{}} \tyR{\md}^{n}}
		{\Gamma \vdash \kdiff{\md}\; \mf{}\; \tm{} :^{\me{}} \tyR{\moda{}}^{n} \to^{\modd{}} \tyR{\mcv{}}^{m}}\\[2mm]
		\inferrule[T-Solve]
		{\Gamma \vdash \mf{} :^{\me{}} \tytuple{\tyR{\mc{}},\tyR{\mcv{}}^{n}} \to^{\modd{}} \tyR{\mcv{}}^{n} \quad \Gamma \vdash \tm{1} :^{\me{}} \tyR{\mcv{}}^{n} \quad \Gamma \vdash \tm{2} :^{\me{}} \tyR{\mc{}} \quad
		\modb{} \leq \mc{}}
		{\Gamma \vdash \ksolve{}\; \mf{}\; \tm{1}\; \tm{2} :^{\me{}} \tyR{\mcv{}}^{n}}
	\end{gather*}
	\caption{Type rules for \cdppl{}.}\label{fig:type-rules}
\end{figure*}

\Cref{fig:type-rules} shows the typing relation of \cdppl{}.
The typing judgment takes the form
\(
\Gamma \vdash \tm{} :^{\me{}} \ty{},
\)
and means that \tm{} can be given type \ty{} in the context \(\Gamma\), as usual.
What is unusual is, like \citet{baudart20}, we annotate the judgment with the \emph{modifier} \me{}, depending on whether the term contains probabilistic constructs.
A random expression will be typed with a modifier \modr{}:
for instance, we have
\(
\vdash \eassume{\mathcal{N}(0,1)}:^{\modr{}} \tyR{\modc{}}.
\)
A deterministic expression will be typed with a modifier \modd{}: for instance \(\vdash \mr{} :^{\modd{}}\tyR{\modc{}} \).
Both these terms have the type \tyR{\modc{}}, meaning we can use them in a differential and non-differential context.
For example, we can supply these terms as arguments to functions with parameters typed \tyR{\mc{}} for any \mc{} (by using \textsc{T-Sub}).
The effect modifiers \me{} decorate not only the typing judgments but also the function arrows (\(\ty{1}\to^{\me{}}\ty{2}\)) to track whether functions represent deterministic or probabilistic computations.
The treatment is strongly inspired by languages with effect systems, such as Links~\cite{lindley12} and Koka~\cite{leijen14}, whose effect rows represent more expressive variants of our modifier \me{}.
The handling of coeffects and the combination of coeffects and effects are inspired by \citet{petricek14} and \citet{gaboardi16}.

The rules \textsc{T-Sub} and \textsc{T-Promote} allow us to coerce between effects and coeffects safely.
\textsc{T-Sub} is a form of weakening: whenever a piece of data is allowed to be used in a non-differentiable context, it can also safely be passed into a differentiable context.
Similarly, it is safe to use a deterministic function in a random context, but not vice-versa.
On the other hand, \textsc{T-Promote} is a form of strengthening, which says that a differentiable piece of data can be used in a non-differentiable context, provided that it was constructed using only non-differentiable components.

The rule \textsc{T-Weaken} allows us to drop unused bindings from the type environment so that \(\mc{} \leq \Gamma\) only restricts the modifiers on types associated with variables we need to type a particular term  (cf. \Cref{fig:meta} for the relevant operations over modifiers, types, and type environments).
This is essential to not make the type system overly conservative.

The rule \textsc{T-Diff} requires that the term we differentiate is a deterministic function. Moreover, the term \kdiff{\md{}} is parametrized by \(\md{}\in\{\moda{},\modb{}\}\), which constrains the parameter type of the differentiated function to analytic or PAP type. This corresponds to our language's two notions of derivatives (analytic and intensional).
The type in the conclusion is attached to the total derivative, which is a \emph{linear function} and, therefore, typed analytic.
The rule \textsc{T-Solve} requires that ODEs are deterministic but not necessarily typed differentiable.
In \Cref{sec:dynamic-semantics}, we discuss smoothness requirements on ODEs.
We restrict \tm{1} to the same type as the type in the conclusion because the solution to an IVP at the initial time is the initial value.
Moreover, we restrict the type of \tm{2} because, as we discuss in \Cref{sec:dynamic-semantics}, \ksolve{} has the meaning of a differentiable ODE integration routine, and it is therefore not sensible to type the IVP solution as analytic w.r.t. time.

The rule \textsc{T-If} is standard with the additions that we restrict the type of the comparand to \(\tyR{\modb{}}\) and propagate effects.
This will ensure that functions whose parameter is involved in the guard of if-then-else terms are only differentiable with \kdiff{\modb{}} (cf. example \eqref{eq:if-example}).

The rules \textsc{T-PrimDist}, \textsc{T-Assume}, \textsc{T-Weight}, \textsc{T-Infer}, and \textsc{T-Dist} give types to the probabilistic constructs of the language.
The \textsc{T-PrimDist} rule constructs \tyname{Dist}-types, which may then be sampled using \kassume{}.
Primitive distributions are typed by \distType{}, defined in \Cref{fig:intrinsic-types}.
We do not allow differentiation of distributions w.r.t. their parameters, which all have the coeffect modifier \modc{}.
Note that \textsc{T-Assume} and \textsc{T-Weight} have modifier \modr{} in the conclusion, which propagates \modr{} from these terms regardless of the effect modifier in the context.
\textsc{T-Infer} expects a probabilistic function (with \modr{}-modifier) returning some \ty{}, and returns a distribution \tydist{\ty{}}.

The rules \textsc{T-Var} and \textsc{T-Abs} are similar to the corresponding rules in the simply typed \(\lambda\)-calculus, with the addition that both are deterministic in the conclusion.
\textsc{T-Var} expects a singleton type environment, which we can always get with \textsc{T-Weaken}, and \textsc{T-Abs} records the effect modifier of the function body in the type.
I.e., its type records any randomness in the abstraction body.
The rule \textsc{T-App} is similar to its corresponding rules in the simply typed \(\lambda\)-calculus, but it also restricts the effects \me{} of the two sub-terms and the arrow type to be the same.

\begin{figure*}
	\begin{subfigure}[b]{0.50\textwidth}
		\begin{align*}
			\intrinsicType(\phi)               & \coloneq (\tyR{\moda{}}, \tyR{\moda{}}, \tyR{\moda{}}) & \phi\in\{+,-,\cdot,\div\} \\
			\intrinsicType(\phi)               & \coloneq (\tyR{\moda{}}, \tyR{\moda{}})                & \phi\in\{\sin,\cos\}      \\
			\intrinsicType(\iwiener{\mpp{}})   & \coloneq (\tyR{\modc{}}, \tyR{\modc{}})                & {}                        \\
			\intrinsicType(\ipdf{\mathcal{N}}) & \coloneq (\tyR{\moda{}}, \tyR{\moda{}}, \tyR{\moda{}}) & {}                        \\
			\intrinsicType(\ipdf{\mathcal{B}}) & \coloneq (\tyR{\modb{}}, \tyR{\modb{}}, \tyR{\modb{}}) & {}
		\end{align*}
	\end{subfigure}%
	\begin{subfigure}[b]{0.50\textwidth}
		\begin{align*}
			\distType(\mDist{}')   & \coloneq (\tyR{\modc{}}, \tyR{\modc{}}, \tyR{\modc{}}) & \mDist{}'\in\{\mathcal{N}, \mathcal{B}\} \\
			\distType(\mathcal{W}) & \coloneq \tyR{\modc{}}\to^{\modd{}}\tyR{\modc{}}       & {}                                       \\
		\end{align*}
	\end{subfigure}
	\caption{Types for intrinsic and distribution symbols}\label{fig:intrinsic-types}
\end{figure*}

The rules \textsc{T-Real}, \textsc{T-PrimApp}, \textsc{T-Tuple}, and \textsc{T-Proj} are standard with the addition that they propagate effect modifiers analog to \textsc{T-App}.
\textsc{T-Real} gives its type the coeffect modifier \modc{} which can always be lowered with \textsc{T-Sub} as needed.

\Cref{fig:intrinsic-types} defines the function \intrinsicType{} which types primitive functions.
In this definition, analytic elementary functions are typed with parameter and return type \tyR{\moda{}}. In contrast, realizations of the Wiener process and the probability density functions are typed with \tyR{\modc{}}, reflecting that these are not differentiable with our \kdiff{\md{}} construct.

Finally, we illustrate the interaction between type rules in the following example:
\begin{gather}
	\inferrule*[Left=T-PrimApp]
	{
		\inferrule*[Left=T-Sub]
		{
			\inferrule*[Left=PrimApp]
			{
				\inferrule*[Left=T-Weaken]
				{
					\inferrule*[Left=T-Promote]
					{
						\inferrule*[Left=PrimApp]
						{\inferrule*[Left=T-Sub]{x:\tyR{\modc{}}\vdash x : \tyR{\modc{}}}{x:\tyR{\modc{}}\vdash x : \tyR{\moda{}}} \quad \inferrule*[right=T-Sub]{x:\tyR{\modc{}}\vdash 1 : \tyR{\modc{}}}{x:\tyR{\modc{}}\vdash 1 : \tyR{\moda{}}}}
						{x:\tyR{\modc{}}\vdash x+1 : \tyR{\moda{}} \quad \modc{} \leq x:\tyR{\modc{}}}
					}
					{x:\tyR{\modc{}}\vdash x+1 : \tyR{\modc{}}}
				}
				{x:\tyR{\modc{}},y:\tyR{\moda{}}\vdash x+1 : \tyR{\modc{}}}
			}
			{x:\tyR{\modc{}},y:\tyR{\moda{}}\vdash \iwiener{\mpp{}}(x+1) : \tyR{\modc{}}}
		}
		{x:\tyR{\modc{}},y:\tyR{\moda{}}\vdash \iwiener{\mpp{}}(x+1) : \tyR{\moda{}}} \quad
		\inferrule*[Right=T-Weaken]
		{y:\tyR{\moda{}}\vdash y:\tyR{\moda{}}}
		{x:\tyR{\modc{}},y:\tyR{\moda{}}\vdash y:\tyR{\moda{}}}
	}
	{x:\tyR{\modc{}},y:\tyR{\moda{}}\vdash \iwiener{\mpp{}}(x+1) + y : \tyR{\moda{}},}
	\label{eq:example-promote}
\end{gather} where, for the sake of brevity we omit effects \me{}, which are all \modd{}; omit the trivial \textsc{T-Var} and \textsc{T-Real}; and omit some premises in \textsc{T-Sub}, \textsc{T-Weaken}, and \textsc{T-PrimApp}.
In~\eqref{eq:example-promote}, we need to promote \(x:\tyR{\modc{}} \vdash x+1 :
\tyR{\moda{}}\) because the term is an argument to \iwiener{\ms{}}.
This, in turn, raises the modifier on \(x:\tyR{\moda{}}\).
With \textsc{T-Weaken}, we prevent the binding \(y:\tyR{\moda{}}\) from also raising its modifier, which allows us to type, e.g.,
\(
\vdash \lambda x:\tyR{\modc{}}.\; \lambda y:\tyR{\moda{}}.\; \iwiener{\mpp{}}(x+1) + y :^{\modd{}} \tyR{\modc{}}\to^{\modd{}}\tyR{\moda{}}\to^{\modd{}}\tyR{\moda{}},
\)
a function typed analytic in its second argument but not the first.

\subsection{Comparison with Other Coeffect Systems}\label{sec:coeffect-compare}

A key feature of our presentation of coeffects, compared to, e.g.,~\citet*{petricek14}, is that we attach coeffect modifiers to the type \tyR{} rather than to arrow types and bindings in the typing environment.  The main reason for this is that the modifier \mc{} only bears meaning for (vectors of) reals.
Attaching modifiers only to the base type \tyR{} also gives us more fine-grained coeffect tracking for tuple types, at the expense of additional annotations.
Consider for example the function
\begin{equation}
	f = \eabs{x:\tyR{}^{2}}{\iwiener{\mpp{}}(\eproj{}{1}{x}) + \eproj{}{2}{x}}.
\end{equation}
A system where only the parameter \(x\) carries a coeffect would have to attach the modifier \modc{} to~\(x\), since it appears as an argument to \iwiener{\mpp{}} (cf. (abs), (app) in Figure 4 in \citet{petricek14}).
Moreover, such a system would also attach \modc{} to \(y\) in \(g = \eabs{y:\tyR{}}{f\; \etuple{0,y}}\), even though  \(y\)  is never passed as an argument to \iwiener{\mpp{}}.
For example, the term
\begin{equation}
	\vdash \ediff{\moda{}}{(\lambda y:\tyR{\moda{}}.\; (\eabs{x:\tytuple{\tyR{\modc{}},\tyR{\moda{}}}}{\iwiener{\mpp{}}(\eproj{}{1}{x}) + \eproj{}{2}{x}})\; \etuple{0,y})}{1}
\end{equation}
is well-typed in our type system, which would not be the case if \(y\) had a single (conservative) modifier.

Compared to \citet{gaboardi16, petricek14}, most explicit occurrences of multiplication can be removed from the typing rules since coeffect multiplication is just the maximum in our system, and \textsc{T-Sub} allows lowering coeffects (and raising effects).
For example, the rule \textsc{T-Promote} in our system directly corresponds to the rule \textsc{pr} in the system of \citet{gaboardi16}, since having an environment \(c \cdot \Gamma\) in the conclusion becomes the same as having a condition \(c \leq \Gamma\) in the premise (and the comonad \(D_r\) in their rule corresponds to multiplication on the result type).
Finally, the distributive laws of \citet{gaboardi16} are trivial in our system since our effects and coeffects have no interesting interaction.

\subsection{Dynamic Semantics}\label{sec:dynamic-semantics}

\begin{figure*}
	\begin{subfigure}[b]{0.40\textwidth}
		\begin{gather*}
			\mw{} \in \dR_{0\leq} \quad % TODO: define in formalization
			\ms{} \in \mathbb{S}\quad
			\tm{} \in \dTerm{} \quad \mv{} \in \dVal{}
		\end{gather*}
		\subcaption{Meta variables}\label{fig:evaluation:meta}
	\end{subfigure}\\
	\begin{subfigure}[b]{0.55\textwidth}
		\begin{gather*}
			\mv{} \Coloneqq \lambda x:\ty{}.\; \tm{} \mid \mr{} \mid \etuple{\mv{1},\ldots,\mv{n}} \mid \mDist(\mv{1},\ldots,\mv{|\mDist|}) \mid \einfer{\mv{}}\qquad
		\end{gather*}
		\subcaption{Values}\label{fig:syntax:values}
	\end{subfigure}\\
	\begin{subfigure}[b]{0.95\textwidth}
		\begin{align*}
			\mE \Coloneqq & \left[\cdot \right] \mid \mE\; \tm{} \mid \mv{}\; \mE \mid \phi(\mv{}, \ldots, \mE, \tm{}, \ldots) \mid \langle\mv{}, \ldots, \mE, \tm{}, \ldots \rangle \mid \eproj{n}{i}\; \mE \mid \eif{\mE}{\tm{1}}{\tm{2}} \\
			\mid\;        & \mDist{}(\mv{}, \ldots, \mE, \tm{}, \ldots) \mid \kassume{}\; \mE \mid \kweight{}\; \mE \mid \kinfer{}\; \mE                                                                                                    \\
			\mid\;        & \kdiff{\md}\; \mE\; \tm{} \mid \kdiff{\md}\; \mv{}\; \mE \mid \ksolve{}\; \mE\; \tm{1}\; \tm{2} \mid \ksolve{}\; \mv{}\; \mE\; \tm{} \mid \ksolve{}\; \mv{1}\; \mv{2}\; \mE
		\end{align*}
		\subcaption{Evaluation contexts.}\label{fig:reduction-rules:context}
	\end{subfigure}
	\begin{subfigure}[b]{0.95\textwidth}
		\begin{gather*}
			\inferrule[E-App]
			{}
			{\mE\left[(\lambda x:\ty{}.\; \tm{})\; \mv{}\right] \rightarrow^{\modd{}} \mE\left[\tm{}\{\mv{} / x\}\right]}\quad
			\inferrule[E-PrimApp]
			{}
			{\mE\left[\phi(\mr{1},\ldots,\mr{|\phi|})\right] \rightarrow^{\modd{}} \mE\left[\den{\phi}(\mr{1},\ldots,\mr{|\phi|})\right]}\\[2mm]
			\inferrule[E-Proj]
			{}
			{\mE\left[\eproj{n}{i}{\langle\mv{1},\ldots,\mv{i},\ldots,\mv{n}\rangle}\right] \rightarrow^{\modd{}} \mE\left[\mv{i}\right]}\\[2mm]
			\inferrule[E-IfTrue]
			{\mr{} > 0}
			{\mE\left[\eif{\mr{}}{\tm{1}}{\tm{2}}\right] \rightarrow^{\modd{}} \mE\left[\tm{1}\right]}\quad
			\inferrule[E-IfFalse]
			{\mr{} \leq 0}
			{\mE\left[\eif{\mr{}}{\tm{1}}{\tm{2}}\right] \rightarrow^{\modd{}} \mE\left[\tm{2}\right]}\\[2mm]
			\inferrule[E-Diff]
			{}
			{\mE\left[\kdiff{\md}\; \mv{1}\; \mv{2}\right] \rightarrow^{\modd{}} \mE\left[\diff{(\mv{1},\mv{2})}\right]}\quad
			\inferrule[E-Solve]
			{}
			{\mE\left[\ksolve{}\; \mv{1}\; \mv{2}\; \mv{3}\right] \rightarrow^{\modd{}} \mE\left[\solve{(\mv{1},\mv{2},\mv{3})}\right]}
		\end{gather*}
		\subcaption{Deterministic reduction rules \(\tm{}\to^{\modd{}}\tm{}'\).}\label{fig:reduction-rules:deterministic}
	\end{subfigure}
	\begin{subfigure}[b]{0.95\textwidth}
		\begin{gather*}
			\inferrule[E-Det]
			{\tm{1}\to^{\modd{}}\tm{2}}
			{(\tm{1},\mw{},\ms{}) \rightarrow^{\modr{}} (\tm{2},\mw{},\ms{})}\quad
			\inferrule[E-Weight]
			{\mr{1} = \max{(\mr{2},0)}}
			{(\mE\left[\kweight{}\; \mr{2}\right],\mw{},\ms{}) \rightarrow^{\modr{}} (\mE\left[\eunit\right],\mr{1}\cdot\mw{},\ms{})}\\[2mm]
			\inferrule[E-AssumeDist]
			{\mr{1} = F^{-1}_{\mDist{}}(\mr{2},\mr{3},\mpp)\quad
				\mDist{} \in \{\mathcal{N},\mathcal{B} \}}
			{(\mE\left[\kassume{}\; \mDist{}(\mr{2},\mr{3})\right],\mw{},\mpp::\ms{}) \rightarrow^{\modr{}} (\mE\left[\mr{1}\right],\mw{},\ms{})}\\[2mm]
			\inferrule[E-AssumeWiener]
			{}
			{(\mE\left[\kassume{}\; \mathcal{W}()\right],\mw{},\mpp{}::\ms{}) \rightarrow^{\modr{}} (\mE\left[\lambda x:\tyR{\modc{}}.\; \iwiener{\mpp{}}(x)\right],\mw{},\ms{})}\\[2mm]
			\inferrule[E-AssumeInfer]
			{F^{-1} =\minfer{(\mv{})}}
			{(\mE\left[\kassume{}\; (\einfer{\mv{}})\right],\mw{},\mpp::\ms{}) \rightarrow^{\modr{}} (\mE\left[F^{-1}(\mpp)\right],\mw{},\ms{})}
		\end{gather*}
		\subcaption{Sampling-based reduction rules \((\tm{},\mw{},\ms{})\to^{\modr{}}(\tm{}',\mw{}',\ms{}')\).}\label{fig:reduction-rules:sampling}
	\end{subfigure}
	\caption{Small step operational semantics of \cdppl{}.}\label{fig:reduction-rules}
\end{figure*}

\Cref{fig:reduction-rules} gives the dynamic \emph{call-by-value} semantics of \cdppl{} in a small-step operational style.
\Cref{fig:evaluation:meta} defines meta variables, where \mw{} are real numbers in the interval \([0,\infty)\) and are, in this section, reserved for likelihood weights; \(s\) are (random) seeds, which are finite sequences of real numbers from the set \(\mathbb{S}=\bigcupplus_{n\in\dN{}}[0,1]^{n}\). These serve as random seeds in the dynamic semantics, and we denote the set of terms as \dTerm{} and the set of values as \dVal{}.
Values are defined in \Cref{fig:syntax:values} and notably include primitive distributions with parameter values and inferred distributions.
In order to use quantile functions over terms we equip closed terms with a total order following \citet{borgstrom16}, based on some total order on \emph{de Bruijn}-encoded terms with real numbers replaced by holes together with the usual dictionary order on \(\dR^n\), where \(n\) is the number of holes in the term.

\Cref{fig:reduction-rules:context} define \emph{evaluation contexts} \mE{}, where the terms of \mE{} includes a unique \emph{hole} \hole{\cdot} and \mE{}\hole{\tm{}} replaces the hole by \tm{} in \mE{}.
We divide the reduction rules into the deterministic reduction relation (\(\rightarrow^{\modd{}}\)), \Cref{fig:reduction-rules:deterministic}, and the sampling-based probabilistic reduction relation (\(\rightarrow^{\modr{}}\)), \Cref{fig:reduction-rules:sampling}.
Our dynamic semantic relies on semantic functions \minfer{}, \diff{}, and \solve{}, which model inference, program differentiation, and IVP solving, respectively.
We call these functions \emph{implementing functions}.
We discuss the properties of these functions as we encounter them in the discussion that follows.

\Cref{fig:reduction-rules:deterministic} defines the deterministic reduction relation \(\tm{} \to^{\modd{}} \tm{}'\).
For convenience, we define some shorthand notation for repeated reductions.
\begin{definition}
	\(\tm{1}\rightarrow^{\modd{}*}\tm{n} = \tm{1}\rightarrow^{\modd{}}\tm{2}\rightarrow^{\modd{}}\cdots\rightarrow^{\modd{}}\tm{n}\) for some \(n\).
\end{definition}

The rules \textsc{E-App}, \textsc{E-If-True}, \textsc{E-If-False}, and \textsc{E-Proj} are standard, where
\(\tm{}\{\mv{}/x\}\) is the substitution of \(x\) with \mv{} in \tm{} after suitable \(\alpha\)-renaming to avoid accidental name capturing. Moreover, we use \den{\cdot} to denote the meaning of terms and primitives as tuples of reals and functions from \(\dR{}^n\) to \(\dR{}^m\).

\textsc{E-PrimApp} reduces primitive function application, where \(\den{\phi}:\dR{}^{|\phi|}\to\dR{}\) is the meaning of the primitive function.
For elementary functions we use their mathematical meaning, where we let calls with undefined argument values (such as division by zero) map to 0 (zero).
\(\den{\ipdf{\mDist{}'}}(\mr{1},\ldots,\mr{|\phi|},\mr{})\) is the probability density function of the primitive distribution \(\mDist{}'\) with parameters \(\mr{1},\ldots,\mr{|\phi|}\) at \mr{}.
The Wiener process generates continuous functions from \([0,\infty)\) to \dR{}.
This set of functions has the same cardinality as \dR{} because of continuity.
We, therefore, assume a function \(g:[0,1]\to(\dR{}\to\dR{})\), which indexes realizations of the (double-sided) Wiener process and define \(\den{\iwiener{\mpp{}}} = g(\mpp{}) \).

The rule \textsc{E-Diff} reduces \kdiff{\md{}} and the function, \(\diff{} : \dVal{} \times \dVal{} \to \dTerm{}\), models differentiation.
Let \(f(x) = \den{\mv{1}}(x)\), where \den{\mv{1}} is the meaning of \mv{1} as a function \(\dR^n\to\dR^m\).
We consider three cases.
If \(f\) is a PAP function,
i.e., it has a PAP representation (definition 4 in \citet*{lee20} or definition II.3 in \citet*{huot23}).
Informally, this means that we can represent \(f\) by a countable family of analytic functions whose domains partition the domain of \(f\).
E.g.
\den{\eabs{x:\tyR{\modb{}}}{\eif{x}{x}{0-x}}} has a possible PAP representation \(\{(\dR_{<0},x \mapsto -x), (\{0\},x \mapsto 0), (\dR_{>0}, x \mapsto -x)\}\), where \(\dR_{<0}\) and \(\dR_{>0}\) are the domains of strictly negative and strictly positive real numbers, respectively.
The meaning of \(\diff{(\mv{1},\mv{2})}\) is then the total derivative \(df_{\den{\mv{2}}} :
\dR^n\to\dR^m\) of \(f\) at \(\den{\mv{2}}\in\dR^n\) in the sense of a total intensional derivative.
If \(f\) is analytic (i.e., we can find a PAP representation with a single element), \(df_{\den{\mv{2}}}\) coincides with the total derivative in the analytic sense.
Otherwise, if \(f\) is not PAP, then \diff{} returns \mv{1}.

\diff{} is naturally implemented with forward-mode AD\footnote{We do not see any difficulty in augmenting our semantics with a differentiation construct suitable for reverse-mode AD.}, but can also be implemented with, e.g., symbolic differentiation or a combination of approaches.
The standard forward-mode AD approach involves lifting elementary functions to operate over so-called dual numbers, where the tangent of the dual number propagates the derivative of a computation via the chain rule \citep{griewank08}.
The \kdiff{\md{}} construct in our language is first-class.
A forward-mode AD implementation, therefore, needs to take care to avoid \emph{pertubation confusion}~\citep{siskind05}, where derivative computations for different invocations of \kdiff{\md{}} accidentally mix and give an incorrect derivative.
We leave the definition of \diff{} open as this work focuses on its composition with the other language constructs, regardless of the term differentiation method.

In the rule \textsc{E-Solve}, the function \(\solve{} : \dVal{} \times \dVal{} \times \dVal{} \to \dTerm{}\) solves an IVP and returns the solution at the given time point as a term.
More precisely, let \(f(x,y) = \den{\mv{1}}(x,y)\), \(y_0 = \den{\mv{2}}\), \(x_1 = \den{\mv{3}}\), and
\begin{equation}
	\label{eq:ivp-def}
	\frac{dy(x)}{dx} = f(x,y(x)) \qquad y(0) = y_0.
\end{equation}
We want \den{\solve{(\mv{1},\mv{2},\mv{3})}} to be \(y(x_1)\), the solution to the IVP~\eqref{eq:ivp-def} at time \(x_1\).

The Picard-Lindel{\"o}f Theorem guarantees the existence of a unique solution to~\eqref{eq:ivp-def} around the initial time \(0\), with the condition that \(f(x,y)\) is continuous and Lipchitz-continuous w.r.t. \(y\) in this interval.
Moreover, if \eqref{eq:ivp-def} has a unique solution \(y(x)\) around \(x_0\) then the derivative of \(y(x;y_0)\) (the solution \(y(x)\) to \eqref{eq:ivp-def} as a function of \(y_0\)) w.r.t. to the components of the initial value \(y_0\) exists, assuming \(f(x,y(x))\) is differentiable w.r.t. to the components of \(y(x)\).
This also extends to the derivative of the solution \(y(x;\theta)\) w.r.t. parameters \(\theta\in\dR^{n_\theta}\) of a parameterized ODE, \(\frac{dy(x)}{dx} = f(x,y(x);\theta)\).
Both \(\frac{dy(x)}{dy_0}\) and \(\frac{dy(x)}{d\theta}\) are themselves solutions to linear IVPs that depend on the Jacobian of \(f\) w.r.t. \(y(x)\) (outlined in \Cref{sec:motivation:sens})~\citep{kelley10}.

Since Lipchitz-continuity follows from analyticity on a closed and bounded interval, we could consider restricting the type of ODE models to
\begin{equation}
	\tytuple{\tyR{\moda{}}, \tyR{\moda{}}^n}\to\tyR{\mcv{}}^n.
	\label{eq:ode-analytic}
\end{equation}
However, the maximal interval for the solution is typically only learned by studying the solution itself.
Moreover, restricting the type of the ODE term in \ksolve{} to \eqref{eq:ode-analytic} is too conservative because there are useful applications where it does not hold (cf. \Cref{sec:motivation:rode}).
If a user wishes, he/she can always restrict the type of the ODE model by wrapping \ksolve{} as:
\begin{equation}
	\lambda z : \tytuple{\tyR{\moda{}}, \tyR{\moda{}}^n}\to^{\modd{}}\tyR{\moda{}}^n.\; \lambda y_0 : \tyR{\moda{}}^n.\; \eabs{x:\tyR{\modb{}}}{\esolve{z}{y_0}{x}}.
\end{equation}

A more pragmatic approach is to implement \(\solve{}(\mv{1},\mv{2},\mv{3}) = \tm{}\) as a term in our language that numerically approximates
\begin{equation}
	\label{eq:ivp-integral-def}
	y(x) = y(0) + \int_{0}^{x} f(\xi,y(\xi))\,d\xi \qquad y(0) = y_0,
\end{equation}
delegating the interpretation of \tm{} as an IVP solution at \(x_1\) to a modeling problem.
We can then use the technique of differentiating the numerical integrator \citep{carmichael97, eberhard99}  to approximate \(\frac{dy}{dy_0}\) and \(\frac{dy}{d\theta}\) as we discuss in \Cref{sec:motivation:sens}.

For example, using a straight-forward Euler discretization, the term
\[
	\mf{}\; (\ldots\mf{}\; (\mf{}\; \mv{2}\; (1 \cdot h))\; (2 \cdot h)\ldots)\; \mv{3}\qquad\text{with } \mf{} = \eabs{y}{\eabs{x}{y + h \cdot \mv{1}\; \etuple{x,y}}},
\]
approximates \eqref{eq:ivp-integral-def} for scalar ODEs, where \(h\) is the numerical integration step size.

\Cref{fig:reduction-rules:sampling} gives the semantics for the probabilistic parts of \cdppl{}.
The semantics are in the weighted sampling-based style of~\citet*{borgstrom16}, where terms are interpreted as functions from a sequence of random numbers to a value and weight.
This view is close to inference methods, and its correctness has been proved for both MCMC~\citep{borgstrom16} and SMC~\citep{lunden21}.
The sampling-based semantics is defined as a relation \((\tm{},\mw{}, \ms{})\to^{\modr{}}(\tm{}',\mw{}', \ms{}')\), between a triplet of terms, weights, and seeds.
As before, we define notation for repeated reductions.
\begin{definition} \label{def:sampling-semantics}
	\((\tm{1},\mw{1},\ms{1})\rightarrow^{\modr{}*}(\tm{n},\mw{n},\ms{n}) = (\tm{1},\mw{1},\ms{1})\rightarrow^{\modr{}}(\tm{2},\mw{2},\ms{2})\rightarrow^{\modr{}}\cdots\rightarrow^{\modr{}}(\tm{n},\mw{n},\ms{n})\) for some \(n\).
\end{definition}

The reduction rule \textsc{E-Det} reduces a term deterministically according to the rules in \Cref{fig:reduction-rules:deterministic}, leaving the weight and the seed unchanged.
\textsc{E-Weight} modifies the weight of the sampling-based relation, where we truncate \mr{2} to ensure that the weight remains positive. Similar to~\citet*{lunden21}, \textsc{E-AssumeDist} is based on the cumulative distribution function \(F_{\mDist{}'}(\mr{2},\mr{3})\) on values with corresponding \emph{quantile function} \(F^{-1}_{\mDist{}'}(\mr{2},\mr{3},\mpp{}) = \inf\;\{\mv{} \mid \mpp{} \le F(\mv{})\}\) for the probability distribution \(\mDist{}'\) with parameters \(\mr{2}\) and \(\mr{3}\), where \(\mpp{} ::
\ms{}\) denotes a seed \ms{} extended with the head \mpp{}.
This rule can easily be extended with additional primitive distributions with real parameters.
Similarly, the rule \textsc{E-AssumeWiener} generates a realization of the Wiener process from the seed head \mpp{} via the measurable family of primitive functions \iwiener{\mpp{}}.

\textsc{E-ExpectationInfer} from \Cref{fig:reduction-rules:deterministic} and \textsc{E-AssumeInfer} construct the inverse cumulative distribution function for the probability distribution that \einfer{\mv{}} represents, via the implementing function
\begin{equation}
	\minfer{} : \dVal{} \to [0,1] \to \dVal{},
\end{equation}
where \(F^{-1} = \minfer{(\mv{}')}\) has the following meaning:
Analogously to~\citet*{borgstrom16} (Section 3.4), we define the result function \oO{\tm{}} and density function \pP{\tm{}} over \(\mathbb{S}\) for a closed term \tm{}.
\begin{definition}
	\begin{align*}
		\oO{\tm{}}(\ms{}) = \begin{cases}
			                    \mv{},    & (\tm{},1,\ms{}) \rightarrow^{\modr{}*} (\mv{},\mw{},[]) \\
			                    \eunit{}, & \text{otherwise}
		                    \end{cases} & \quad
		\pP{\tm{}}(\ms{}) = \begin{cases}
			                    \mw{}, & (\tm{},1,\ms{}) \rightarrow^{\modr{}*} (\mv{},\mw{},[]) \\
			                    0,     & \text{otherwise},
		                    \end{cases}
	\end{align*}
\end{definition}
\noindent
where \([]\) denotes the empty sequence.
By Lemma 9 in~\citet*{borgstrom16}, these functions are measurable for the basic probabilistic calculus.
With \(\tm{} = \mv{}\; \eunit{}\), we then form the un-normalized cumulative distribution function,
\begin{equation}\label{eq:infer-F}
	F'(\mv{}) = \int \pP{\tm{}}(\ms{})\cdot\left[\oO{\tm{}}(\ms{})\in\{\mv{}' \mid \mv{}'\leq \mv{}\}\right]d\ms{},
\end{equation}
over values.  Like \citet*{borgstrom16} we use a simple total order on values \(\mv{}\) composed from some order on program terms with holes for real numbers and the dictionary order on tuples of real numbers with the same arity as the number of holes of the term.
We define \(N_t = \int \pP{\tm{}}(s)\,ds\). The normalized cumulative distribution function then becomes \(F(v)=\frac{F'(\mv{})}{N_t}\) if \(N_t \neq 0\) or \(F(\mv{}) = 1\) otherwise.
The function \minfer{} models an inference algorithm that approximates \eqref{eq:infer-F}, for instance using Monte-Carlo methods~\citep*{borgstrom16, lunden21}.
The implementation of inference, \Cref{sec:implementation}, which is not part of our contribution, includes both SMC and MCMC algorithms.
Even though the semantics of our language allow arbitrary nesting of \kinfer{}, supporting such an implementation is non-trivial (cf. \citet{rainforth18}), and therefore, nesting \kinfer{} in our implementation is not supported.

\textsc{E-AssumeInfer} then, similar to \textsc{E-AssumeDist}, samples from the distribution encoded by \(F^{-1}\) by applying it to the head of the seed.

\section{Mechanized Semantics and Type Safety}\label{sec:mechanization}

This section presents the main theorems and summarizes the proofs from the Agda formalization\footnote{\url{https://github.com/miking-lang/dppl-formalization}} of the semantics in \Cref{sec:formalization}, based on the formalization of locally nameless sets by \citet{pitts23}.

We consider top-level terms in a deterministic context, with typing judgments \(\vdash \tm{} :^{\modd{}} \ty{}\).
We want well-typed terms to be safe, in the sense of the standard type-safety guarantee that we can always reduce closed, well-typed terms to values.
Moreover, we want deterministically typed terms (i.e., typed with the \(\Gamma \vdash \tm{} :^{\modd{}} \ty{}\) relation) to be independent of the random seed and weight when executed in a probabilistic context.

In the formalization, the semantics is parameterized on total functions for \diff{}, \solve{}, and \minfer{} for differentiation, IVP solving, and sampling from inferred distributions.
The terms returned from these functions are assumed to have the appropriate type in any type environment where the arguments to the functions do.

As is common, we prove type safety via theorems of progress and preservation.
Progress guarantees that a well-typed term \tm{} is either a value or can be reduced one step, and preservation guarantees that reductions are type-preserving.
The deterministic reduction relation \(\tm{} \rightarrow^{\modd{}} \tm{}'\) (\Cref{fig:reduction-rules:deterministic}) internally depends on the sample-based reduction relation \((\tm{},\mw{},\ms{})\to^{\modr{}}(\tm{}',\mw{}',\ms{}')\) (\Cref{fig:reduction-rules:sampling}) in the rule \textsc{E-Infer}.
Hence, we also prove progress and preservation for terms typed \(\vdash \tm{} :^{\modr{}} \ty{}\).
The theorems follow.

\begin{theorem}[Progress \(\modd{}\)]
	\label{thm:det-prog}
	Suppose \(\vdash \tm{} :^{\modd{}} \ty{}\) holds. Then either \tm{} is a value or else there is some \(\tm{}'\) such that \(\tm{} \rightarrow^{\modd{}} \tm{}'\).
\end{theorem}

\begin{theorem}[Preservation \(\modd{}\)]
	If\; \(\vdash \tm{} :^{\me{}} \ty{}\) and \(\tm{} \rightarrow^{\modd{}} \tm{}'\), then \(\vdash \tm{}' :^{\me{}} \ty{}\).
\end{theorem}

\begin{theorem}[Progress \(\modr{}\)]
	Suppose \(\vdash \tm{} :^{\me{}} \ty{}\) holds. Then either \tm{} is a value or else for any \mw{}, \mpp{}, and \ms{}, there are \(\tm{}',\mw{}',\) and \(\ms{}'\) such that \((\tm{},\mw{},\mpp{}::\ms{}) \rightarrow^{\modr{}}(\tm{}',\mw{}',\ms{}')\), where \mpp{} in \(\mpp{}::\ms{}\) denotes the head of a seed and \ms{} its remainder.
\end{theorem}

\begin{theorem}[Preservation \(\modr{}\)]
	If\; \(\vdash \tm{} :^{\me{}} \ty{}\) and \((\tm{},\mw{},\ms{}) \rightarrow^{\modr{}}(\tm{}',\mw{}',\ms{}')\), then \(\vdash \tm{}' :^{\me{}} \ty{}\).
\end{theorem}

We are now ready to prove our main theorem on type safety.

\begin{theorem}[Type Safety]
	If\; \(\vdash \tm{} :^{\modd{}} \ty{}\) and \(\tm{} \rightarrow^{\modd{}*} \tm{}'\) for some irreducible \(\tm{}'\), then \(\tm{}'\) is a value.
\end{theorem}

To prove that deterministically typed terms are not influenced by probabilistic effects, we first show that deterministic reduction of deterministically typed terms \(\Gamma \vdash \tm{} :^{\modd{}} \ty{}\) is indeed deterministic.

\begin{theorem}[Determinism \(\modd{}\)]
	\label{thm:det-det}
	If\; \(\tm{}\rightarrow^{\modd{}}\tm{1}\) and \(\tm{}\rightarrow^{\modd{}}\tm{2},\) then \(\tm{1} = \tm{2}\).
\end{theorem}

Similarly, given a seed, a term typed \(\Gamma \vdash \tm{} :^{\modr{}} \ty{}\) will reduce deterministically.

\begin{theorem}[Determinism \(\modr{}\)]
	\label{thm:det-rnd}
	If\; \((\tm{},\mw{},\ms{})\rightarrow^{\modr{}}(\tm{1},\mw{1},\ms{1})\), and\\ \((\tm{},\mw{},\ms{})\rightarrow^{\modr{}}(\tm{2},\mw{2},\ms{2}),\) then \(\tm{1} = \tm{2}\land\mw{1}=\mw{2}\land\ms{1}=\ms{2}\).
\end{theorem}

As a corollary to Theorems \ref{thm:det-prog} and \ref{thm:det-rnd}, a term typed \(\vdash \tm{} :^{\modd{}} \ty{}\) will never change the seed or weight in the \(\rightarrow^{\modr{}}\) relation.
This holds since \tm{} will either be a value or have a transition that does not affect the weight and seed (using rule \textsc{E-Det}), and by determinism, any other transition must give the same result.

\begin{corollary}
	If\; \(\vdash \tm{} :^{\modd{}} \ty{}\) holds, then for all \(\mw{},\ms{},\tm{}',\mw{}'\), and \(\ms{}'\) such that \((\tm{},\mw{},\ms{})\rightarrow^{\modr{}*}(\tm{}',\mw{}',\ms{}')\), we have \(\mw{}=\mw{}'\land\ms{}=\ms{}'\).
\end{corollary}

\section{Implementation}\label{sec:implementation}

We implement a prototype of \Cdppl{} as an extended subset of the compiler for the universal probabilistic programming language \miking{} \citep{Broman:2019} \cppl{}\footnote{\url{https://github.com/miking-lang/miking-dppl}} \citep{lunden22}
.
In particular, \cdppl{} extends \cppl{} with \kdiff{}, \ksolve{}, and \(\mathcal{W}\), as well as a type-checker that implements the static semantics of \Cref{sec:static-semantics}.
\cppl{} does not implement an explicit \ipdf{\mDist{}'} primitive but instead includes an \kobserve{} term.
\Cdppl{} is implemented in roughly 2500 lines of code.

We implement \diff{} based on dynamically tagged dual numbers, which allows for a correct first-class derivative construct in a higher-order functional language \citep{siskind08}. Moreover, we also implement the ODE integrators underlying \solve{} as differentiable terms in the \cdppl{} implementation. Currently, it includes Euler-forward and explicit fourth-order Runge-Kutta ODE integrators.
We base the Wiener process implementation on code\footnote{\url{https://github.com/lazyppl-team/lazyppl/blob/89f162641a27e2db3c8632f9ae32c33b78db5ea2/src/WienerDemo.lhs\#L70}} from \lazyppl{} \citep{paquet21}.

The type checker implements \textsc{T-Sub} using joins/meets (cf. \citep{pierce02subtyping}) and implements the promotion rule \textsc{T-Promote} in three steps as follows:
(i) for each term \tm{} and type environment \(\Gamma\), apply \textsc{T-Weaken} until \(\dom{}(\Gamma) = FV(\tm{})\); (ii) find the maximum \mc{} s.t. \(\mc{} \leq \Gamma\) for the weakened \(\Gamma\); and (iii) use \textsc{T-Promote} with the maximum \mc{} from (ii) to promote the type of \tm{}.
For example, when type-checking the following term, \(\eabs{x:\tyR{\moda{}}}{1 + 1}\), it weakens \(\Gamma\) to the empty environment \(\{\}\) when type-checking \(1+1\), which allows it to promote the body to \tyR{\modc{}} which is less restrictive than \tyR{\moda{}}, the return type of \((+)\). The type of the whole term is \(\tyR{\moda{}}\to^{\modd{}}\tyR{\modc{}}\).

\section{Case Studies}\label{sec:case-studies}

This section presents three case studies (A, B, and C in \Cref{fig:overview}) on the expressiveness of \cdppl{}.
All case studies are implemented as programs in the implementation described in \Cref{sec:implementation}.

\subsection{A: Two Methods of ODE Sensitivities}
\label{sec:motivation:sens}

\begin{figure*}
	\includegraphics[width=\textwidth]{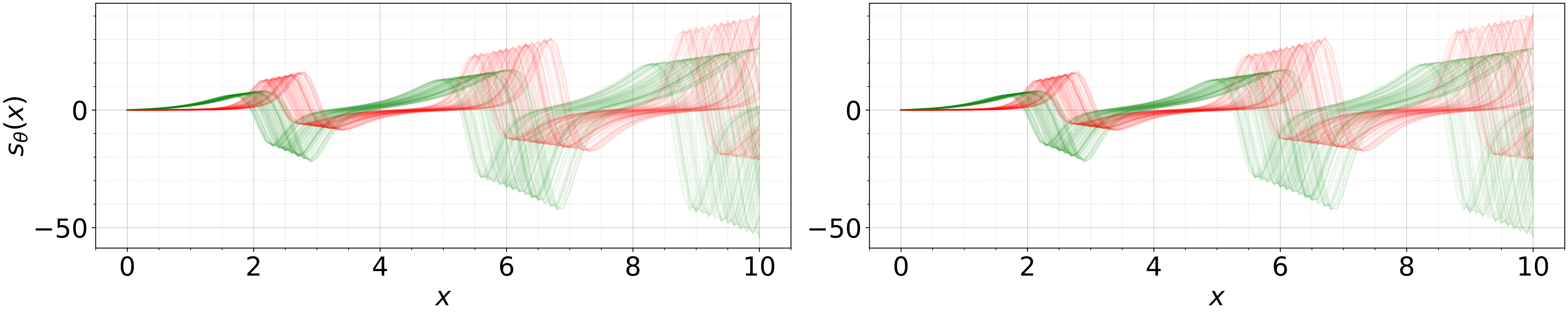}
	\caption{Sensitivities of the Lotka-Volterra model for the prey (green) and predator (red) densities w.r.t. the prey per-capita growth rate parameter, using \kdiff{} of \ksolve{} (Equation \eqref{eq:diff-ode-sens-example} extended to \(\tyR{}^n\)) (left) and \ksolve{} with sensitivity equations using \kdiff{} (Equation \eqref{eq:ode-diff-sens-example} extended to \(\tyR{}^n\)) (right), for different values of the parameter. The horizontal axis denotes time, and the integration used Runge-Kutta with step size $10^{-3}$.}
	\label{fig:sens-example}
\end{figure*}

Gradient-based optimization techniques, such as gradient descent, require the gradient of the objective function. Suppose the objective function, in turn, solves an IVP parameterized on the variables of the objective function. In that case, the gradient of the objective function includes the derivative of the solution of the ODE w.r.t. its parameters. These are known as the sensitivities of the ODE.
A parameterized ODE model generally has \(n_y\) states and \(n_\theta\) parameters.
For now, assume the scalar case \(n_y = n_\theta = 1\), and we discuss the general case at the end of this example.
The composition of \kdiff{} and \ksolve{} allows us to express the sensitivities, \(s_{\theta}(x) = \frac{dy(x)}{d\theta}\), w.r.t. to a single parameter \(\theta \in \dR\) in two ways depending on whether we nest \kdiff{} in \ksolve{} or vice versa.

For both cases, assume a smooth parameterized scalar model \(\idname{ODE} : \tyR{\moda{}} \to \tytuple{\tyR{\moda{}},\tyR{\moda{}}} \to \tytuple{\tyR{\moda{}}} \) with the initial value \(y_0\).
We can express \(s_\theta(x)\) by differentiating \ksolve{} as
\begin{equation}
	\label{eq:diff-ode-sens-example}
	\lambda \theta : \tyR{\moda{}}.\; \eabs{x : \tyR{\modb{}}}{\ediffone{\moda{}}{(\eabs{\theta : \tyR{\moda{}}}{\esolve{({\idname{ODE}\; \theta)}}{y_0}{x}})}{\theta}}.
\end{equation}
We may also compute the ODE sensitives by augmenting the ODE with an additional equation.
Differentiation of both sides of the ODE equation w.r.t. \(\theta\) gives
\begin{equation}
	\label{eq:ode-sens-eq}
	\frac{d}{d\theta}\frac{dy}{dx} = \frac{d}{d\theta}f\Longleftrightarrow \frac{d}{dx}\frac{dy}{d\theta} = \frac{df}{dy}\frac{dy}{d\theta} + \frac{df}{d\theta} \Longleftrightarrow \frac{ds}{dx} = \frac{df}{dy}s + \frac{df}{d\theta},
\end{equation}
where \(y=y(x)\), \(s=s(x)\), \(f=f(x, y(x);\theta)\),
and \(\frac{df}{d\theta}\) should be read as the derivative of \(f({}\cdot{},{}\cdot{};\theta)\) w.r.t. to its third argument \(\theta\).
Hence, we can augment our ODE with~\eqref{eq:ode-sens-eq} to compute the sensitivity. We can, therefore, also express the sensitivities as
\begin{equation}
	\label{eq:ode-diff-sens-example}
	\begin{array}{ll}
		\lambda \theta : \tyR{\moda{}}.\; \lambda x : \tyR{\modb{}}. & \klet\; z = \lambda \etuple{x, \etuple{y, s}}.                                                                                                                                                     \\
		{}                                                           & \quad \elet{\idname{ODE}}{\lambda \theta : \tyR{\moda{}}.\; \eabs{y : \tyR{\moda{}}}{\idname{ODE}\; \theta\; \etuple{x,y}}}                                                                        \\
		{}                                                           & \quad \kin\; \etuple{\idname{ODE}\; \theta\; y, (\ediffone{\moda{}}{(\idname{ODE}\; \theta)}{y}) \cdot s + \ediffone{\moda{}}{(\eabs{\theta : \tyR{\moda{}}}{\idname{ODE}\; \theta\; y})}{\theta}} \\
		{}                                                           & \kin\; \eletin{\etuple{y,s}}{\esolve{z}{\etuple{y_0,0}}{x}}\; s,
	\end{array}
\end{equation}
where we extend the original parameterized model, \idname{ODE}, with the sensitivity model and project the sensitivities from the extended IVP solution.
The benefit of~\eqref{eq:ode-diff-sens-example} is that we can get the sensitivities without requiring that \ksolve{} is differentiable, which lets us implement it using some external solver library.
The idea of differentiating the ODE solver to compute sensitivities, Equation \eqref{eq:diff-ode-sens-example}, is not new~\citep{carmichael97, eberhard99}, but it demonstrates the expressiveness of a language with both \kdiff{} and \ksolve{}.

\Cref{fig:sens-example} shows the sensitivities of an example ODE for the two different methods for computing sensitivities in our implementation as generalized to non-scalar ODEs (\(n_y > 1\)).

\subsection{B: Random ODEs}
\label{sec:motivation:rode}

\begin{figure*}[t]
	\includegraphics[width=\textwidth]{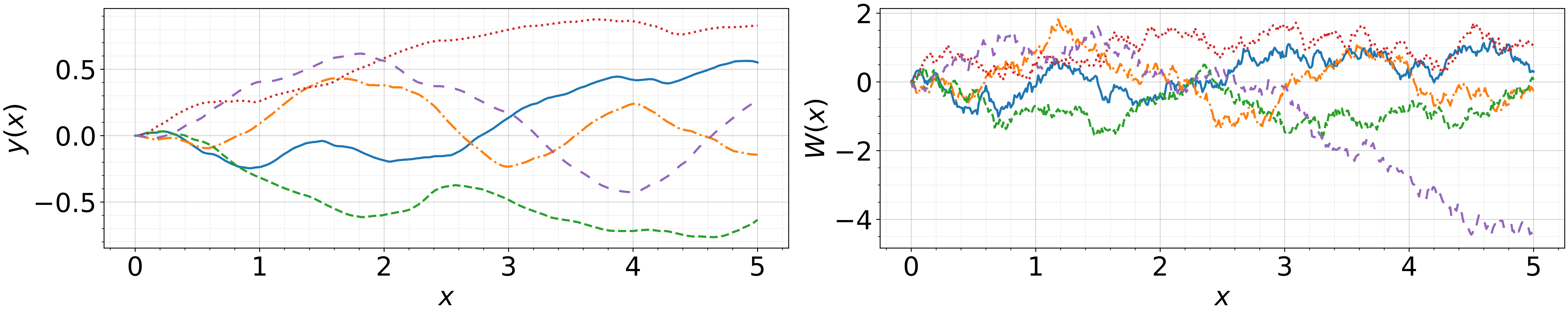}
	\caption{Samples from~\eqref{eq:rode-program-example1} (left) and their Wiener realizations (right).
		The horizontal axis denotes time and integration used Euler forward with a step size $10^{-4}$.
	}
	\label{fig:rode-example}
\end{figure*}

The composition of PPL constructs and \ksolve{} allows us to conveniently express RODEs.
RODE models are useful on their own, and there is a conjugacy between RODEs and Stochastic Differential Equation (SDE) models, which have been studied for several non-trivial models~\citep{neckel17, imkeller01}.
For example, we can encode a solution trace, in the form of an n-tuple of reals for the RODE in example 2.3 in \citet{jentzen11} (ch. 2), encoding a synthetic system, as the following probabilistic term:
\begin{equation}
	\begin{aligned}
		 & \klet\; \idname{RODE} = \lambda \_ : \tyunit{}.                                                                                                                                            \\
		 & \quad \begin{aligned}
			          & \elet{w}{\eassume{\mathcal{W}()}}\; \kin{}\; \eletin{z}{\eabs{\etuple{x,y} : \tytuple{\tyR{\modc{}}, \tyR{\moda{}}}}{\sin(w\; x)-y}} \\
			          & \etuple{\ksolve{}\; z\; y_0\; (1\cdot h), \ksolve{}\; z\; y_0\; (2\cdot h), \ldots, \ksolve{}\; z\; y_0\; (n\cdot h)}
		         \end{aligned} \\
		 & \kin\; \kinfer\; \idname{RODE},
	\end{aligned}
	\label{eq:rode-program-example1}
\end{equation}
with initial values: \(y_0:\tyR{\mc{}}\) and solution trace interval \([h,nh]\).
The model \eqref{eq:rode-program-example1} is simplified for the sake of presentation, compared to the implementation, which for numerical stability reasons instead folds \ksolve{} over \(n\) time intervals of size \(h\).
\Cref{fig:rode-example} depicts sample traces from the resulting distribution.

Although the RODE \eqref{eq:rode-program-example1} is quite simple, it demonstrates that it is natural to express RODEs by composing \ksolve{}, \kinfer{}, and the Wiener process.

\subsection{C: Bayesian Parameter Estimation and ODE Sensitivities}
\label{sec:motivation:parameter-estimation}

As we discussed in \Cref{sec:motivation1}, the composition of \ksolve{} and the probabilistic constructs allows us to express parameter estimation problems in a Bayesian setting.
This class of problems forms an integral part of the field of system identification, where dynamical models are fitted to data \citep{Ljung1999}.
The parameter estimation problem for non-linear state space models consists of an ODE, \(\frac{dy(x)}{dx} = f(x, y(x);\theta)\), with some unknown parameters \(\theta\in\dR{}^{n_{\theta}}\) that models a dynamic system, an output function \(g(y(x;\theta))\in\dR{}^{n_g}\) that model what we can observe in our system, and noisy measurements from the modeled system (\(y(x;\theta)\) is the solution, \(y(x)\), to the ODE parametrized on \(\theta\)).
For example, \(g\) can model the relation between invisible internal states \(y(x)\) of the system model and sensor outputs that we can observe.

The example in \Cref{fig:big-example} expresses a parameter estimation problem in addition to expressing a probabilistic model for the sensitivity of \(y(x;\theta)\) w.r.t. \(\theta\).
In particular, with the Lotka-Volterra model (\Cref{apdx:lotak-volterra}) with \(\theta \in \dR{}\) as the prey per-capita growth rate (the remaining parameters are fixed), \(y_1(x;\theta)\) and \(y_2(x;\theta)\) as prey and predator densities, respectively, then \( g(y(x;\theta)) = y_1(x;\theta)\).
The code snippet in \Cref{fig:big-example} then expresses the posterior distribution for a trace of \(\frac{d}{d\theta}y(x;\theta)\) over \(x\), with priors \(\theta \sim \mathcal{N}(1,1)\), \(\sigma \sim \mathcal{B}(2,2)\); and for a data set of input-output pairs \((x, o)\), \(o \sim \mathcal{N}(g(y(x,\theta)), \sigma)\).
The data used for this example consisted of 11 data points sampled with a frequency of $10$ from \(g(y(x;\theta))\) with \(\theta=1.5\) and with additive zero mean Gaussian noise with variance \(0.04\).
Integration used Runge-Kutta with a step size \(10^{-3}\).

In summary, examples A to C illustrate how the ability to compose \kdiff{}, \ksolve{}, and probabilistic constructs allows us to naturally express and combine models from different domains and condition them on data.

\section{Related Work}\label{sec:related}

\subsection{Differentiable Programming, and IVP Solutions}

\citet{carmichael97} and \citet{eberhard99} study differentiation of ODE integration routines to compute ODE sensitivities.
More recently, \citet{ma21} compare the performance of traditional methods for computing sensitivities to solver differentiation in the \julia{} language.
Neither of these works studies the composition of IVP solutions and differentiation in a formal language context.

From a more fundamental mathematical perspective, \citet{gofen09} studies the relation between AD and solutions to IVPs. The author discusses a wider definition of elementary functions, from the standard set of tabulated functions to solutions of a certain class of ODEs.
Because AD is driven by the differentiation of elementary functions, with this extended notion of elementary functions, IVP solving constitutes a method for AD, while at the same time, AD constitutes a method for solving IVPs via Taylor integration. Although our work is more programming language-oriented, these results illustrated interesting aspects of the composition of IVP solving and program differentiation.

As demonstrated by \citet{bangaru21}, languages that include an integration primitive \citep{bangaru21, sherman21} can express solutions to separable ODEs. In contrast, the IVP \ksolve{} primitive allows the solution of both separable and non-separable ODE models.

\probzelus{}~\citep{baudart20} extends the hybrid synchronous language \zelus{}~\citep{benveniste18}, which extends synchronous languages with declarative ODE models, with PPL constructs.
Their formalization does not explicitly include ODE models. Their static semantics separate deterministic and non-deterministic computations by decorating the typing judgment and arrow types like \cdppl{} does with the effect modifier \me{}.
\probzelus{} does not include program differentiation.

The PPL framework \pymc{} \citep{abril23} and the PPL \stan{} \citep{carpenter17} explicitly include an IVP solver primitive. In addition, \pymc{} supports differentiation through a tensor library based on \theano{} \citep{al16}.
Naturally, these works are implementation-focused, while our work studies the idea of differentiable probabilistic programs in a formal language setting.
In contrast to \stan{}, our type system tracks different notions of differentiability and randomness, which allows it to reject a larger class of ill-defined programs at compile time.
\pymc{} is a framework built in Python and, therefore, lacks a static type system.

\subsection{Differentiable Programming and Probabilistic Programming Languages}

\citet{borgstrom16} give an operational semantics for an untyped \(\lambda\)-calculus extended with probabilistic language constructs, forming a universal PPL, and we base our PPL semantics on this work.
In the traditional presentation of PPLs (as exemplified by Borgström), each program encodes a measure over result values; in other words, the whole program is probabilistic. A more recent line of work \cite{heunen17,scibior18,vakar19,matache22,dash23} describes how probability measures form a monad over quasi-Borel spaces \cite{heunen17}, which gives a natural model of computation where deterministic and probabilistic programs can co-exist.

There are several works on the correctness of AD (e.g., \cite{mazza21, abadiplotkin19, huot23, lee20, krawiec22, elliott18, barthe20}) and a large body of more implementation focused works on AD (e.g., \citep{siskind08, wang19, shaikhha19, pearlmutter08} and \citet*{baydin2018automatic} gives a survey for machine learning applications).
Our work directly uses the notion of intensional derivatives introduced by \citet{lee20}.

\citet{huot23} build upon both these lines of work, extending quasi-Borel spaces into \(\omega\)PAP spaces to support differentiation, investigating semantic properties such as the differentiability of the density functions of probabilistic programs. \citet{lew23} go in a slightly different direction, considering automatic differentiation of expected values of probabilistic programs using a similar semantic foundation. Their type system distinguishes between smooth and non-smooth types, and the language includes a coercion operation from non-smooth to smooth types, while our language handles this coercion in the type system.
The smoothness tracking of \citet{lew23} is also present in \citet{becker24} which includes a type rule to implicitly convert non-smooth types to smooth types similar to our \textsc{T-Sub} with \(\ty{} = \tyR{\modb{}}\) and \(\ty{}'=\tyR{\moda{}}\).
\citet*{lee23} propose a static smoothness analysis for a formal imperative PPL based on abstract interpretation with applications toward variational inference.
\citet*{barthe20} demonstrates the power of open logical relations by proving continuity properties for a language with if-then-else constructs and refinement types.
We instead track smoothness via a coeffect-based type system and plan to base a future proof on the differentiability properties of our language on the technique of Barthe et al.
Neither of these works considers a first-class differentiation operator as part of the PPL syntax.

\citet{sherman21} propose a higher-order differentiable programming language, \(\lambda_s\), with a first-class derivative, root finding, optimization, and integration primitives.
The integration primitive, defined over the compact domain \([0,1]\), allows \(\lambda_s\) to express certain probability distributions.
\(\lambda_s\) generalized the notion of derivatives to \emph{Clarke} derivatives \citep{clarke75} while our type system instead distinguishes between different notions of derivatives.
\citet{michel24} extends program differentiation to correctly differentiate parametric discontinuities in a first-order language.

\subsection{Effects and Coeffects}

Languages with effect systems, such as Koka \cite{leijen14} or Links \cite{lindley12}, allow users to write direct-style code with multiple effects, implicitly representing a monadic computation.
More generally, the idea of using monads to model alternative notions of computation (or effects) has a long history in the field (see, e.g., \citet*{moggi91}).  These type systems can statically enforce a separation between effectful and pure code.

Complementary to effects is the notion of coeffects that allows the static reasoning about how function parameters are used in a program \cite{petricek14, brunel14}.
\citet{gaboardi16} develops a calculus for combining effects and coeffects via effect-graded monads, and the type system of \cdppl{} is directly inspired by these works.

\section{Conclusion}\label{sec:conclusion}

This paper studies safe composition in differentiable probabilistic programming, which combines the domains of differentiable programming, ordinary differential equations, and probabilistic programming languages.
It demonstrates the concept of differentiable probabilistic programming on a series of examples and
develops formal semantics for a core differentiable probabilistic programming language.
In particular, the static semantics track randomness through an effect-based system and different notions of differentiability through a coeffect-based system.
This ensures the type system can reject ill-defined derivatives and ordinary differential equation models.
Moreover, it provides mechanized type soundness proofs and mechanized proofs of the static separation between deterministic and random terms.

One avenue for increased expressiveness is to extend \ksolve{} to Differential-Algebraic Equations (DAEs), which are implicit ODEs with additional algebraic constraints.
DAEs allow physical modeling from first principles and are the fundamental representation of equation-based modeling languages~\citep{broman10}, but solving them typically requires systematic differentiation of their individual equations~\citep{petzold82}.
It seems possible to combine our type system with known DAE structure analysis algorithms~\citep{pantelides88,pryce01} to statically reject certain ill-formed DAEs combined with PPL constructs. However, further research work is needed to study this in a formal setting.

Connecting the static and dynamic semantics, it remains to be proved that well-typed functions with parameter type \(\tyR{\moda{}}^n\) and parameter type \(\tyR{\modb{}}^n\) are indeed analytic and PAP functions, respectively. More research is needed on a suitable domain for this proof, that also admits the semantic implementing functions \minfer{}, \diff{}, and \solve{}.

Another direction of future work is coeffect and effect modifier inference.  The sub-typing relation in our type system only considers these modifiers. Since they are related by simple inequalities, this would produce constraints similar to those for existing type inference algorithms for linear types~\citep{matsuda20}.
Coupled with appropriate error messages, this extension would spare the user details in the type system while retaining its static guarantees.

%%
%% The acknowledgments section is defined using the "acks" environment
%% (and NOT an unnumbered section). This ensures the proper
%% identification of the section in the article metadata, and the
%% consistent spelling of the heading.
\begin{acks}
	The authors thank the reviewers for their valuable feedback. This work was financially supported by the Swedish Research Council (Grant No. 2018-04329).  The second author was supported by the Swedish Research Council through the projects Accelerating Managed Languages (Grant No. 2020-05346) and Pervasive Memory Safety through Ownership Types (Grant No. 2023-05526).
\end{acks}

%%
%% The next two lines define the bibliography style to be used, and
%% the bibliography file.
\bibliographystyle{ACM-Reference-Format}
\bibliography{references}

%%% -*-BibTeX-*-
%%% Do NOT edit. File created by BibTeX with style
%%% ACM-Reference-Format-Journals [18-Jan-2012].

\begin{thebibliography}{69}

%%% ====================================================================
%%% NOTE TO THE USER: you can override these defaults by providing
%%% customized versions of any of these macros before the \bibliography
%%% command.  Each of them MUST provide its own final punctuation,
%%% except for \shownote{}, \showDOI{}, and \showURL{}.  The latter two
%%% do not use final punctuation, in order to avoid confusing it with
%%% the Web address.
%%%
%%% To suppress output of a particular field, define its macro to expand
%%% to an empty string, or better, \unskip, like this:
%%%
%%% \newcommand{\showDOI}[1]{\unskip}   % LaTeX syntax
%%%
%%% \def \showDOI #1{\unskip}           % plain TeX syntax
%%%
%%% ====================================================================

\ifx \showCODEN    \undefined \def \showCODEN     #1{\unskip}     \fi
\ifx \showDOI      \undefined \def \showDOI       #1{#1}\fi
\ifx \showISBNx    \undefined \def \showISBNx     #1{\unskip}     \fi
\ifx \showISBNxiii \undefined \def \showISBNxiii  #1{\unskip}     \fi
\ifx \showISSN     \undefined \def \showISSN      #1{\unskip}     \fi
\ifx \showLCCN     \undefined \def \showLCCN      #1{\unskip}     \fi
\ifx \shownote     \undefined \def \shownote      #1{#1}          \fi
\ifx \showarticletitle \undefined \def \showarticletitle #1{#1}   \fi
\ifx \showURL      \undefined \def \showURL       {\relax}        \fi
% The following commands are used for tagged output and should be
% invisible to TeX
\providecommand\bibfield[2]{#2}
\providecommand\bibinfo[2]{#2}
\providecommand\natexlab[1]{#1}
\providecommand\showeprint[2][]{arXiv:#2}

\bibitem[Abadi and Plotkin(2019)]%
        {abadiplotkin19}
\bibfield{author}{\bibinfo{person}{Mart\'{\i}n Abadi} {and}
  \bibinfo{person}{Gordon~D. Plotkin}.} \bibinfo{year}{2019}\natexlab{}.
\newblock \showarticletitle{A simple differentiable programming language}.
\newblock \bibinfo{journal}{\emph{Proceedings of the ACM on Programming
  Languages}} \bibinfo{volume}{4}, \bibinfo{number}{POPL}, Article
  \bibinfo{articleno}{38} (\bibinfo{date}{dec} \bibinfo{year}{2019}),
  \bibinfo{numpages}{28}~pages.
\newblock
\urldef\tempurl%
\url{https://doi.org/10.1145/3371106}
\showDOI{\tempurl}


\bibitem[Abril-Pla et~al\mbox{.}(2023)]%
        {abril23}
\bibfield{author}{\bibinfo{person}{Oriol Abril-Pla}, \bibinfo{person}{Virgile
  Andreani}, \bibinfo{person}{Colin Carroll}, \bibinfo{person}{Larry Dong},
  \bibinfo{person}{Christopher~J. Fonnesbeck}, \bibinfo{person}{Maxim
  Kochurov}, \bibinfo{person}{Ravin Kumar}, \bibinfo{person}{Junpeng Lao},
  \bibinfo{person}{Christian~C. Luhmann}, \bibinfo{person}{Osvaldo~A. Martin},
  \bibinfo{person}{Michael Osthege}, \bibinfo{person}{Ricardo Vieira},
  \bibinfo{person}{Thomas Wiecki}, {and} \bibinfo{person}{Robert Zinkov}.}
  \bibinfo{year}{2023}\natexlab{}.
\newblock \showarticletitle{{PyMC}: a modern, and comprehensive probabilistic
  programming framework in {Python}}.
\newblock \bibinfo{journal}{\emph{PeerJ Computer Science}}  \bibinfo{volume}{9}
  (\bibinfo{date}{Sept.} \bibinfo{year}{2023}), \bibinfo{pages}{e1516}.
\newblock
\showISSN{2376-5992}
\urldef\tempurl%
\url{https://doi.org/10.7717/peerj-cs.1516}
\showDOI{\tempurl}


\bibitem[Al-Rfou et~al\mbox{.}(2016)]%
        {al16}
\bibfield{author}{\bibinfo{person}{Rami Al-Rfou}, \bibinfo{person}{Guillaume
  Alain}, \bibinfo{person}{Amjad Almahairi}, \bibinfo{person}{Christof
  Angermueller}, \bibinfo{person}{Dzmitry Bahdanau}, \bibinfo{person}{Nicolas
  Ballas}, \bibinfo{person}{Fr{\'e}d{\'e}ric Bastien}, \bibinfo{person}{Justin
  Bayer}, \bibinfo{person}{Anatoly Belikov}, \bibinfo{person}{Alexander
  Belopolsky}, {et~al\mbox{.}}} \bibinfo{year}{2016}\natexlab{}.
\newblock \showarticletitle{Theano: A Python framework for fast computation of
  mathematical expressions}.
\newblock \bibinfo{journal}{\emph{arXiv e-prints}} (\bibinfo{year}{2016}).
\newblock
\showeprint{https://doi.org/10.48550/arXiv.1605.02688}


\bibitem[Bangaru et~al\mbox{.}(2021)]%
        {bangaru21}
\bibfield{author}{\bibinfo{person}{Sai~Praveen Bangaru}, \bibinfo{person}{Jesse
  Michel}, \bibinfo{person}{Kevin Mu}, \bibinfo{person}{Gilbert Bernstein},
  \bibinfo{person}{Tzu-Mao Li}, {and} \bibinfo{person}{Jonathan Ragan-Kelley}.}
  \bibinfo{year}{2021}\natexlab{}.
\newblock \showarticletitle{Systematically differentiating parametric
  discontinuities}.
\newblock \bibinfo{journal}{\emph{ACM Transactions on Graphics}}
  \bibinfo{volume}{40}, \bibinfo{number}{4} (\bibinfo{date}{Aug.}
  \bibinfo{year}{2021}), \bibinfo{pages}{1--18}.
\newblock
\showISSN{0730-0301, 1557-7368}
\urldef\tempurl%
\url{https://doi.org/10.1145/3450626.3459775}
\showDOI{\tempurl}


\bibitem[Barthe et~al\mbox{.}(2020)]%
        {barthe20}
\bibfield{author}{\bibinfo{person}{Gilles Barthe}, \bibinfo{person}{Raphaëlle
  Crubillé}, \bibinfo{person}{Ugo~Dal Lago}, {and} \bibinfo{person}{Francesco
  Gavazzo}.} \bibinfo{year}{2020}\natexlab{}.
\newblock \showarticletitle{On the {Versatility} of {Open} {Logical}
  {Relations}}. In \bibinfo{booktitle}{\emph{Programming {Languages} and
  {Systems}}}, \bibfield{editor}{\bibinfo{person}{Peter Müller}} (Ed.).
  \bibinfo{publisher}{Springer International Publishing},
  \bibinfo{address}{Cham}, \bibinfo{pages}{56--83}.
\newblock
\showISBNx{978-3-030-44914-8}
\urldef\tempurl%
\url{https://doi.org/10.1007/978-3-030-44914-8_3}
\showDOI{\tempurl}


\bibitem[Baudart et~al\mbox{.}(2020)]%
        {baudart20}
\bibfield{author}{\bibinfo{person}{Guillaume Baudart}, \bibinfo{person}{Louis
  Mandel}, \bibinfo{person}{Eric Atkinson}, \bibinfo{person}{Benjamin Sherman},
  \bibinfo{person}{Marc Pouzet}, {and} \bibinfo{person}{Michael Carbin}.}
  \bibinfo{year}{2020}\natexlab{}.
\newblock \showarticletitle{Reactive probabilistic programming}. In
  \bibinfo{booktitle}{\emph{Proceedings of the 41st {ACM} {SIGPLAN}
  {Conference} on {Programming} {Language} {Design} and {Implementation}}}
  \emph{(\bibinfo{series}{{PLDI} 2020})}. \bibinfo{publisher}{Association for
  Computing Machinery}, \bibinfo{address}{New York, NY, USA},
  \bibinfo{pages}{898--912}.
\newblock
\showISBNx{978-1-4503-7613-6}
\urldef\tempurl%
\url{https://doi.org/10.1145/3385412.3386009}
\showDOI{\tempurl}


\bibitem[Baydin et~al\mbox{.}(2018)]%
        {baydin2018automatic}
\bibfield{author}{\bibinfo{person}{Atilim~Gunes Baydin},
  \bibinfo{person}{Barak~A Pearlmutter}, \bibinfo{person}{Alexey~Andreyevich
  Radul}, {and} \bibinfo{person}{Jeffrey~Mark Siskind}.}
  \bibinfo{year}{2018}\natexlab{}.
\newblock \showarticletitle{{Automatic differentiation in machine learning: a
  survey}}.
\newblock \bibinfo{journal}{\emph{{Journal of machine learning research}}}
  \bibinfo{volume}{18}, \bibinfo{number}{153} (\bibinfo{year}{2018}),
  \bibinfo{pages}{1--43}.
\newblock
\urldef\tempurl%
\url{http://jmlr.org/papers/v18/17-468.html}
\showURL{%
\tempurl}


\bibitem[Becker et~al\mbox{.}(2024)]%
        {becker24}
\bibfield{author}{\bibinfo{person}{McCoy~R. Becker},
  \bibinfo{person}{Alexander~K. Lew}, \bibinfo{person}{Xiaoyan Wang},
  \bibinfo{person}{Matin Ghavami}, \bibinfo{person}{Mathieu Huot},
  \bibinfo{person}{Martin~C. Rinard}, {and} \bibinfo{person}{Vikash~K.
  Mansinghka}.} \bibinfo{year}{2024}\natexlab{}.
\newblock \showarticletitle{Probabilistic {Programming} with {Programmable}
  {Variational} {Inference}}.
\newblock \bibinfo{journal}{\emph{Reproduction Packager for Article
  "Probabilistic Programming with Programmable Variational Inference"}}
  \bibinfo{volume}{8}, \bibinfo{number}{PLDI} (\bibinfo{date}{June}
  \bibinfo{year}{2024}), \bibinfo{pages}{233:2123--233:2147}.
\newblock
\urldef\tempurl%
\url{https://doi.org/10.1145/3656463}
\showDOI{\tempurl}


\bibitem[Benveniste et~al\mbox{.}(2018)]%
        {benveniste18}
\bibfield{author}{\bibinfo{person}{Albert Benveniste}, \bibinfo{person}{Timothy
  Bourke}, \bibinfo{person}{Beno\^{\i}t Caillaud}, \bibinfo{person}{Jean-Louis
  Cola\c{c}o}, \bibinfo{person}{C\'edric Pasteur}, {and} \bibinfo{person}{Marc
  Pouzet}.} \bibinfo{year}{2018}\natexlab{}.
\newblock \showarticletitle{{Building a Hybrid Systems Modeler on Synchronous
  Languages Principles}}.
\newblock \bibinfo{journal}{\emph{Proc. IEEE}} (\bibinfo{year}{2018}).
\newblock
\urldef\tempurl%
\url{https://www.di.ens.fr/~pouzet/bib/ieee18.pdf}
\showURL{%
\tempurl}


\bibitem[Bingham et~al\mbox{.}(2019)]%
        {bingham19}
\bibfield{author}{\bibinfo{person}{Eli Bingham}, \bibinfo{person}{Jonathan~P.
  Chen}, \bibinfo{person}{Martin Jankowiak}, \bibinfo{person}{Fritz Obermeyer},
  \bibinfo{person}{Neeraj Pradhan}, \bibinfo{person}{Theofanis Karaletsos},
  \bibinfo{person}{Rohit Singh}, \bibinfo{person}{Paul Szerlip},
  \bibinfo{person}{Paul Horsfall}, {and} \bibinfo{person}{Noah~D. Goodman}.}
  \bibinfo{year}{2019}\natexlab{}.
\newblock \showarticletitle{Pyro: deep universal probabilistic programming}.
\newblock \bibinfo{journal}{\emph{Journal of Machine Learning Research}}
  \bibinfo{volume}{20}, \bibinfo{number}{1} (\bibinfo{date}{jan}
  \bibinfo{year}{2019}), \bibinfo{pages}{973–978}.
\newblock
\showISSN{1532-4435}
\urldef\tempurl%
\url{http://jmlr.org/papers/v20/18-403.html}
\showURL{%
\tempurl}


\bibitem[Borgstr\"{o}m et~al\mbox{.}(2016)]%
        {borgstrom16}
\bibfield{author}{\bibinfo{person}{Johannes Borgstr\"{o}m},
  \bibinfo{person}{Ugo Dal~Lago}, \bibinfo{person}{Andrew~D. Gordon}, {and}
  \bibinfo{person}{Marcin Szymczak}.} \bibinfo{year}{2016}\natexlab{}.
\newblock \showarticletitle{A lambda-calculus foundation for universal
  probabilistic programming}.
\newblock \bibinfo{journal}{\emph{ACM SIGPLAN Notices}} \bibinfo{volume}{51},
  \bibinfo{number}{9} (\bibinfo{date}{sep} \bibinfo{year}{2016}),
  \bibinfo{pages}{33–46}.
\newblock
\showISSN{0362-1340}
\urldef\tempurl%
\url{https://doi.org/10.1145/3022670.2951942}
\showDOI{\tempurl}


\bibitem[Broman(2010)]%
        {broman10}
\bibfield{author}{\bibinfo{person}{David Broman}.}
  \bibinfo{year}{2010}\natexlab{}.
\newblock \emph{\bibinfo{title}{{Meta-Languages and Semantics for
  Equation-Based Modeling and Simulation}}}.
\newblock \bibinfo{thesistype}{Ph.\,D. Dissertation}.
  \bibinfo{school}{{Department of Computer and Information Science,
  Link{\"o}ping University}}, \bibinfo{address}{Sweden}.
\newblock


\bibitem[Broman(2019)]%
        {Broman:2019}
\bibfield{author}{\bibinfo{person}{David Broman}.}
  \bibinfo{year}{2019}\natexlab{}.
\newblock \showarticletitle{{A Vision of Miking: Interactive Programmatic
  Modeling, Sound Language Composition, and Self-Learning Compilation}}. In
  \bibinfo{booktitle}{\emph{Proceedings of the 12th ACM SIGPLAN International
  Conference on Software Language Engineering}} \emph{(\bibinfo{series}{SLE
  '19})}. ACM, \bibinfo{pages}{55--60}.
\newblock
\urldef\tempurl%
\url{https://doi.org/10.1145/3357766.3359531}
\showDOI{\tempurl}


\bibitem[Brunel et~al\mbox{.}(2014)]%
        {brunel14}
\bibfield{author}{\bibinfo{person}{Alo\"{\i}s Brunel}, \bibinfo{person}{Marco
  Gaboardi}, \bibinfo{person}{Damiano Mazza}, {and} \bibinfo{person}{Steve
  Zdancewic}.} \bibinfo{year}{2014}\natexlab{}.
\newblock \showarticletitle{A Core Quantitative Coeffect Calculus}. In
  \bibinfo{booktitle}{\emph{Proceedings of the 23rd European Symposium on
  Programming Languages and Systems - Volume 8410}}.
  \bibinfo{publisher}{Springer-Verlag}, \bibinfo{address}{Berlin, Heidelberg},
  \bibinfo{pages}{351–370}.
\newblock
\showISBNx{9783642548321}
\urldef\tempurl%
\url{https://doi.org/10.1007/978-3-642-54833-8_19}
\showDOI{\tempurl}


\bibitem[Carmichael et~al\mbox{.}(1997)]%
        {carmichael97}
\bibfield{author}{\bibinfo{person}{Gregory~R. Carmichael},
  \bibinfo{person}{Adrian Sandu}, {and} \bibinfo{person}{florian~A. Potra}.}
  \bibinfo{year}{1997}\natexlab{}.
\newblock \showarticletitle{Sensitivity analysis for atmospheric chemistry
  models via automatic differentiation}.
\newblock \bibinfo{journal}{\emph{Atmospheric Environment}}
  \bibinfo{volume}{31}, \bibinfo{number}{3} (\bibinfo{date}{Feb.}
  \bibinfo{year}{1997}), \bibinfo{pages}{475--489}.
\newblock
\showISSN{1352-2310}
\urldef\tempurl%
\url{https://doi.org/10.1016/S1352-2310(96)00168-9}
\showDOI{\tempurl}


\bibitem[Carpenter et~al\mbox{.}(2017)]%
        {carpenter17}
\bibfield{author}{\bibinfo{person}{Bob Carpenter}, \bibinfo{person}{Andrew
  Gelman}, \bibinfo{person}{Matthew~D Hoffman}, \bibinfo{person}{Daniel Lee},
  \bibinfo{person}{Ben Goodrich}, \bibinfo{person}{Michael Betancourt},
  \bibinfo{person}{Marcus~A Brubaker}, \bibinfo{person}{Jiqiang Guo},
  \bibinfo{person}{Peter Li}, {and} \bibinfo{person}{Allen Riddell}.}
  \bibinfo{year}{2017}\natexlab{}.
\newblock \showarticletitle{Stan: A probabilistic programming language}.
\newblock \bibinfo{journal}{\emph{Journal of statistical software}}
  \bibinfo{volume}{76} (\bibinfo{year}{2017}).
\newblock
\urldef\tempurl%
\url{https://doi.org/10.18637/jss.v076.i01}
\showDOI{\tempurl}


\bibitem[Clarke(1975)]%
        {clarke75}
\bibfield{author}{\bibinfo{person}{Frank~H. Clarke}.}
  \bibinfo{year}{1975}\natexlab{}.
\newblock \showarticletitle{Generalized gradients and applications}.
\newblock \bibinfo{journal}{\emph{{Transactions of the American Mathematical
  Society}}}  \bibinfo{volume}{205} (\bibinfo{year}{1975}),
  \bibinfo{pages}{247–247}.
\newblock
\showISSN{0002-9947}
\urldef\tempurl%
\url{https://doi.org/10.1090/s0002-9947-1975-0367131-6}
\showDOI{\tempurl}


\bibitem[Cusumano-Towner et~al\mbox{.}(2019)]%
        {cusumano-towner19}
\bibfield{author}{\bibinfo{person}{Marco~F. Cusumano-Towner},
  \bibinfo{person}{Feras~A. Saad}, \bibinfo{person}{Alexander~K. Lew}, {and}
  \bibinfo{person}{Vikash~K. Mansinghka}.} \bibinfo{year}{2019}\natexlab{}.
\newblock \showarticletitle{Gen: a general-purpose probabilistic programming
  system with programmable inference}. In \bibinfo{booktitle}{\emph{Proceedings
  of the 40th ACM SIGPLAN Conference on Programming Language Design and
  Implementation}} (Phoenix, AZ, USA) \emph{(\bibinfo{series}{PLDI 2019})}.
  \bibinfo{publisher}{Association for Computing Machinery},
  \bibinfo{address}{New York, NY, USA}, \bibinfo{pages}{221–236}.
\newblock
\showISBNx{9781450367127}
\urldef\tempurl%
\url{https://doi.org/10.1145/3314221.3314642}
\showDOI{\tempurl}


\bibitem[Dash et~al\mbox{.}(2023)]%
        {dash23}
\bibfield{author}{\bibinfo{person}{Swaraj Dash}, \bibinfo{person}{Younesse
  Kaddar}, \bibinfo{person}{Hugo Paquet}, {and} \bibinfo{person}{Sam Staton}.}
  \bibinfo{year}{2023}\natexlab{}.
\newblock \showarticletitle{Affine Monads and Lazy Structures for Bayesian
  Programming}.
\newblock \bibinfo{journal}{\emph{Proceedings of the ACM on Programming
  Languages}} \bibinfo{volume}{7}, \bibinfo{number}{{POPL}}
  (\bibinfo{year}{2023}), \bibinfo{pages}{1338--1368}.
\newblock
\urldef\tempurl%
\url{https://doi.org/10.1145/3571239}
\showDOI{\tempurl}


\bibitem[Eberhard and Bischof(1999)]%
        {eberhard99}
\bibfield{author}{\bibinfo{person}{Peter Eberhard} {and}
  \bibinfo{person}{Christian Bischof}.} \bibinfo{year}{1999}\natexlab{}.
\newblock \showarticletitle{Automatic differentiation of numerical integration
  algorithms}.
\newblock \bibinfo{journal}{\emph{Math. Comp.}} \bibinfo{volume}{68},
  \bibinfo{number}{226} (\bibinfo{date}{April} \bibinfo{year}{1999}),
  \bibinfo{pages}{717--732}.
\newblock
\showISSN{0025-5718}
\urldef\tempurl%
\url{https://doi.org/10.1090/S0025-5718-99-01027-3}
\showDOI{\tempurl}


\bibitem[Elliott(2018)]%
        {elliott18}
\bibfield{author}{\bibinfo{person}{Conal Elliott}.}
  \bibinfo{year}{2018}\natexlab{}.
\newblock \showarticletitle{The simple essence of automatic differentiation}.
\newblock \bibinfo{journal}{\emph{Proceedings of the ACM on Programming
  Languages}} \bibinfo{volume}{2}, \bibinfo{number}{ICFP} (\bibinfo{date}{July}
  \bibinfo{year}{2018}), \bibinfo{pages}{70:1--70:29}.
\newblock
\urldef\tempurl%
\url{https://doi.org/10.1145/3236765}
\showDOI{\tempurl}


\bibitem[Fritzson(2014)]%
        {fritzson2014principles}
\bibfield{author}{\bibinfo{person}{Peter Fritzson}.}
  \bibinfo{year}{2014}\natexlab{}.
\newblock \bibinfo{booktitle}{\emph{{Principles of object-oriented modeling and
  simulation with Modelica 3.3: a cyber-physical approach}}}.
\newblock \bibinfo{publisher}{{John Wiley \& Sons}}.
\newblock


\bibitem[Gaboardi et~al\mbox{.}(2016)]%
        {gaboardi16}
\bibfield{author}{\bibinfo{person}{Marco Gaboardi}, \bibinfo{person}{Shin-ya
  Katsumata}, \bibinfo{person}{Dominic Orchard}, \bibinfo{person}{Flavien
  Breuvart}, {and} \bibinfo{person}{Tarmo Uustalu}.}
  \bibinfo{year}{2016}\natexlab{}.
\newblock \showarticletitle{Combining effects and coeffects via grading}.
\newblock \bibinfo{journal}{\emph{ACM SIGPLAN Notices}} \bibinfo{volume}{51},
  \bibinfo{number}{9} (\bibinfo{date}{Sept.} \bibinfo{year}{2016}),
  \bibinfo{pages}{476--489}.
\newblock
\showISSN{0362-1340}
\urldef\tempurl%
\url{https://doi.org/10.1145/3022670.2951939}
\showDOI{\tempurl}


\bibitem[Ghahramani(2015)]%
        {ghahramani2015probabilistic}
\bibfield{author}{\bibinfo{person}{Zoubin Ghahramani}.}
  \bibinfo{year}{2015}\natexlab{}.
\newblock \showarticletitle{{Probabilistic machine learning and artificial
  intelligence}}.
\newblock \bibinfo{journal}{\emph{Nature}} \bibinfo{volume}{521},
  \bibinfo{number}{7553} (\bibinfo{year}{2015}), \bibinfo{pages}{452--459}.
\newblock
\urldef\tempurl%
\url{https://doi.org/10.1038/nature14541}
\showDOI{\tempurl}


\bibitem[Gofen(2009)]%
        {gofen09}
\bibfield{author}{\bibinfo{person}{Alexander~M. Gofen}.}
  \bibinfo{year}{2009}\natexlab{}.
\newblock \showarticletitle{The ordinary differential equations and automatic
  differentiation unified}.
\newblock \bibinfo{journal}{\emph{Complex Variables and Elliptic Equations}}
  \bibinfo{volume}{54}, \bibinfo{number}{9} (\bibinfo{year}{2009}),
  \bibinfo{pages}{825--854}.
\newblock
\urldef\tempurl%
\url{https://doi.org/10.1080/17476930902998852}
\showDOI{\tempurl}


\bibitem[Goodman et~al\mbox{.}(2008)]%
        {goodman08}
\bibfield{author}{\bibinfo{person}{Noah~D. Goodman}, \bibinfo{person}{Vikash~K.
  Mansinghka}, \bibinfo{person}{Daniel Roy}, \bibinfo{person}{Keith Bonawitz},
  {and} \bibinfo{person}{Joshua~B. Tenenbaum}.}
  \bibinfo{year}{2008}\natexlab{}.
\newblock \showarticletitle{Church: a language for generative models}. In
  \bibinfo{booktitle}{\emph{Proceedings of the Twenty-Fourth Conference on
  Uncertainty in Artificial Intelligence}} (Helsinki, Finland)
  \emph{(\bibinfo{series}{UAI'08})}. \bibinfo{publisher}{AUAI Press},
  \bibinfo{address}{Arlington, Virginia, USA}, \bibinfo{pages}{220–229}.
\newblock
\showISBNx{0974903949}


\bibitem[Goodman and Stuhlm\"{u}ller(2014)]%
        {goodman14}
\bibfield{author}{\bibinfo{person}{Noah~D Goodman} {and}
  \bibinfo{person}{Andreas Stuhlm\"{u}ller}.} \bibinfo{year}{2014}\natexlab{}.
\newblock \bibinfo{title}{{The Design and Implementation of Probabilistic
  Programming Languages}}.
\newblock \bibinfo{howpublished}{\url{http://dippl.org}}.
\newblock
\newblock
\shownote{Accessed: 2024-2-15}.


\bibitem[Griewank and Walther(2008)]%
        {griewank08}
\bibfield{author}{\bibinfo{person}{Andreas Griewank} {and}
  \bibinfo{person}{Andrea Walther}.} \bibinfo{year}{2008}\natexlab{}.
\newblock \bibinfo{booktitle}{\emph{Evaluating derivatives: principles and
  techniques of algorithmic differentiation}}.
\newblock \bibinfo{publisher}{SIAM, Philadelphia, PA}.
\newblock


\bibitem[Heunen et~al\mbox{.}(2017)]%
        {heunen17}
\bibfield{author}{\bibinfo{person}{Chris Heunen}, \bibinfo{person}{Ohad
  Kammar}, \bibinfo{person}{Sam Staton}, {and} \bibinfo{person}{Hongseok
  Yang}.} \bibinfo{year}{2017}\natexlab{}.
\newblock \showarticletitle{A convenient category for higher-order probability
  theory}. In \bibinfo{booktitle}{\emph{32nd Annual {ACM/IEEE} Symposium on
  Logic in Computer Science, {LICS} 2017, Reykjavik, Iceland, June 20-23,
  2017}}. \bibinfo{publisher}{{IEEE} Computer Society}, \bibinfo{pages}{1--12}.
\newblock
\urldef\tempurl%
\url{https://doi.org/10.1109/LICS.2017.8005137}
\showDOI{\tempurl}


\bibitem[Huot et~al\mbox{.}(2023)]%
        {huot23}
\bibfield{author}{\bibinfo{person}{Mathieu Huot}, \bibinfo{person}{Alexander~K.
  Lew}, \bibinfo{person}{Vikash~K. Mansinghka}, {and} \bibinfo{person}{Sam
  Staton}.} \bibinfo{year}{2023}\natexlab{}.
\newblock \showarticletitle{{\(\omega\)}PAP Spaces: Reasoning Denotationally
  About Higher-Order, Recursive Probabilistic and Differentiable Programs}. In
  \bibinfo{booktitle}{\emph{{LICS}}}. \bibinfo{pages}{1--14}.
\newblock
\urldef\tempurl%
\url{https://doi.org/10.1109/LICS56636.2023.10175739}
\showDOI{\tempurl}


\bibitem[Imkeller and Schmalfuss(2001)]%
        {imkeller01}
\bibfield{author}{\bibinfo{person}{Peter Imkeller} {and}
  \bibinfo{person}{Björn Schmalfuss}.} \bibinfo{year}{2001}\natexlab{}.
\newblock \showarticletitle{The {Conjugacy} of {Stochastic} and {Random}
  {Differential} {Equations} and the {Existence} of {Global} {Attractors}}.
\newblock \bibinfo{journal}{\emph{Journal of Dynamics and Differential
  Equations}} \bibinfo{volume}{13}, \bibinfo{number}{2} (\bibinfo{date}{April}
  \bibinfo{year}{2001}), \bibinfo{pages}{215--249}.
\newblock
\showISSN{1572-9222}
\urldef\tempurl%
\url{https://doi.org/10.1023/A:1016673307045}
\showDOI{\tempurl}


\bibitem[Jentzen and Kloeden(2011)]%
        {jentzen11}
\bibfield{author}{\bibinfo{person}{Arnulf Jentzen} {and}
  \bibinfo{person}{Peter~E. Kloeden}.} \bibinfo{year}{2011}\natexlab{}.
\newblock \bibinfo{booktitle}{\emph{Taylor Approximations for Stochastic
  Partial Differential Equations}}.
\newblock \bibinfo{publisher}{Society for Industrial and Applied Mathematics}.
\newblock
\urldef\tempurl%
\url{https://doi.org/10.1137/1.9781611972016}
\showDOI{\tempurl}
\showeprint{https://epubs.siam.org/doi/pdf/10.1137/1.9781611972016}


\bibitem[Kelley and Peterson(2010)]%
        {kelley10}
\bibfield{author}{\bibinfo{person}{Walter~G. Kelley} {and}
  \bibinfo{person}{Allan~C. Peterson}.} \bibinfo{year}{2010}\natexlab{}.
\newblock \bibinfo{booktitle}{\emph{The {Theory} of {Differential} {Equations}:
  {Classical} and {Qualitative}}}.
\newblock \bibinfo{publisher}{Springer New York}, \bibinfo{address}{New York,
  NY}.
\newblock
\showISBNx{978-1-4419-5782-5 978-1-4419-5783-2}
\urldef\tempurl%
\url{https://doi.org/10.1007/978-1-4419-5783-2}
\showDOI{\tempurl}


\bibitem[Krawiec et~al\mbox{.}(2022)]%
        {krawiec22}
\bibfield{author}{\bibinfo{person}{Faustyna Krawiec}, \bibinfo{person}{Simon
  Peyton~Jones}, \bibinfo{person}{Neel Krishnaswami}, \bibinfo{person}{Tom
  Ellis}, \bibinfo{person}{Richard~A. Eisenberg}, {and} \bibinfo{person}{Andrew
  Fitzgibbon}.} \bibinfo{year}{2022}\natexlab{}.
\newblock \showarticletitle{Provably correct, asymptotically efficient,
  higher-order reverse-mode automatic differentiation}.
\newblock \bibinfo{journal}{\emph{Proceedings of the ACM on Programming
  Languages}} \bibinfo{volume}{6}, \bibinfo{number}{POPL} (\bibinfo{date}{Jan.}
  \bibinfo{year}{2022}), \bibinfo{pages}{1--30}.
\newblock
\showISSN{2475-1421}
\urldef\tempurl%
\url{https://doi.org/10.1145/3498710}
\showDOI{\tempurl}


\bibitem[Lee et~al\mbox{.}(2023)]%
        {lee23}
\bibfield{author}{\bibinfo{person}{Wonyeol Lee}, \bibinfo{person}{Xavier
  Rival}, {and} \bibinfo{person}{Hongseok Yang}.}
  \bibinfo{year}{2023}\natexlab{}.
\newblock \showarticletitle{Smoothness {Analysis} for {Probabilistic}
  {Programs} with {Application} to {Optimised} {Variational} {Inference}}.
\newblock \bibinfo{journal}{\emph{Artifact for the Paper "Smoothness Analysis
  for Probabilistic Programs with Application to Optimised Variational
  Inference"}} \bibinfo{volume}{7}, \bibinfo{number}{POPL}
  (\bibinfo{date}{Jan.} \bibinfo{year}{2023}), \bibinfo{pages}{12:335--12:366}.
\newblock
\urldef\tempurl%
\url{https://doi.org/10.1145/3571205}
\showDOI{\tempurl}


\bibitem[Lee et~al\mbox{.}(2020)]%
        {lee20}
\bibfield{author}{\bibinfo{person}{Wonyeol Lee}, \bibinfo{person}{Hangyeol Yu},
  \bibinfo{person}{Xavier Rival}, {and} \bibinfo{person}{Hongseok Yang}.}
  \bibinfo{year}{2020}\natexlab{}.
\newblock \showarticletitle{On Correctness of Automatic Differentiation for
  Non-Differentiable Functions}. In \bibinfo{booktitle}{\emph{Advances in
  Neural Information Processing Systems 33: Annual Conference on Neural
  Information Processing Systems 2020, NeurIPS 2020, December 6-12, 2020,
  virtual}}, \bibfield{editor}{\bibinfo{person}{Hugo Larochelle},
  \bibinfo{person}{Marc'Aurelio Ranzato}, \bibinfo{person}{Raia Hadsell},
  \bibinfo{person}{Maria{-}Florina Balcan}, {and} \bibinfo{person}{Hsuan{-}Tien
  Lin}} (Eds.).
\newblock
\urldef\tempurl%
\url{https://proceedings.neurips.cc/paper/2020/hash/4aaa76178f8567e05c8e8295c96171d8-Abstract.html}
\showURL{%
\tempurl}


\bibitem[Leijen(2014)]%
        {leijen14}
\bibfield{author}{\bibinfo{person}{Daan Leijen}.}
  \bibinfo{year}{2014}\natexlab{}.
\newblock \showarticletitle{Koka: Programming with Row Polymorphic Effect
  Types}. In \bibinfo{booktitle}{\emph{Proceedings 5th Workshop on
  Mathematically Structured Functional Programming, MSFP@ETAPS 2014, Grenoble,
  France, 12 April 2014}} \emph{(\bibinfo{series}{{EPTCS}},
  Vol.~\bibinfo{volume}{153})}, \bibfield{editor}{\bibinfo{person}{Paul~Blain
  Levy} {and} \bibinfo{person}{Neel Krishnaswami}} (Eds.).
  \bibinfo{pages}{100--126}.
\newblock
\urldef\tempurl%
\url{https://doi.org/10.4204/EPTCS.153.8}
\showDOI{\tempurl}


\bibitem[Lew et~al\mbox{.}(2023)]%
        {lew23}
\bibfield{author}{\bibinfo{person}{Alexander~K. Lew}, \bibinfo{person}{Mathieu
  Huot}, \bibinfo{person}{Sam Staton}, {and} \bibinfo{person}{Vikash~K.
  Mansinghka}.} \bibinfo{year}{2023}\natexlab{}.
\newblock \showarticletitle{{ADEV:} Sound Automatic Differentiation of Expected
  Values of Probabilistic Programs}.
\newblock \bibinfo{journal}{\emph{Proceedings of the ACM on Programming
  Languages}} \bibinfo{volume}{7}, \bibinfo{number}{{POPL}}
  (\bibinfo{year}{2023}), \bibinfo{pages}{121--153}.
\newblock
\urldef\tempurl%
\url{https://doi.org/10.1145/3571198}
\showDOI{\tempurl}


\bibitem[Lindley and Cheney(2012)]%
        {lindley12}
\bibfield{author}{\bibinfo{person}{Sam Lindley} {and} \bibinfo{person}{James
  Cheney}.} \bibinfo{year}{2012}\natexlab{}.
\newblock \showarticletitle{Row-based effect types for database integration}.
  In \bibinfo{booktitle}{\emph{Proceedings of the 8th {ACM} {SIGPLAN} Workshop
  on Types in Languages Design and Implementation, {TLDI} 2012, Philadelphia,
  PA, USA, Saturday, January 28, 2012}},
  \bibfield{editor}{\bibinfo{person}{Benjamin~C. Pierce}} (Ed.).
  \bibinfo{publisher}{{ACM}}, \bibinfo{pages}{91--102}.
\newblock
\urldef\tempurl%
\url{https://doi.org/10.1145/2103786.2103798}
\showDOI{\tempurl}


\bibitem[Ljung(1999)]%
        {Ljung1999}
\bibfield{author}{\bibinfo{person}{L. Ljung}.} \bibinfo{year}{1999}\natexlab{}.
\newblock \bibinfo{booktitle}{\emph{System Identification: Theory for the User}
  (\bibinfo{edition}{2nd} ed.)}.
\newblock \bibinfo{publisher}{Prentice Hall}.
\newblock
\showISBNx{0-13-656695-2}


\bibitem[Lucassen and Gifford(1988)]%
        {lucassen88}
\bibfield{author}{\bibinfo{person}{J.~M. Lucassen} {and} \bibinfo{person}{D.~K.
  Gifford}.} \bibinfo{year}{1988}\natexlab{}.
\newblock \showarticletitle{Polymorphic effect systems}. In
  \bibinfo{booktitle}{\emph{Proceedings of the 15th ACM SIGPLAN-SIGACT
  Symposium on Principles of Programming Languages}} (San Diego, California,
  USA) \emph{(\bibinfo{series}{POPL '88})}. \bibinfo{publisher}{Association for
  Computing Machinery}, \bibinfo{address}{New York, NY, USA},
  \bibinfo{pages}{47–57}.
\newblock
\showISBNx{0897912527}
\urldef\tempurl%
\url{https://doi.org/10.1145/73560.73564}
\showDOI{\tempurl}


\bibitem[Lundén et~al\mbox{.}(2021)]%
        {lunden21}
\bibfield{author}{\bibinfo{person}{Daniel Lundén}, \bibinfo{person}{Johannes
  Borgström}, {and} \bibinfo{person}{David Broman}.}
  \bibinfo{year}{2021}\natexlab{}.
\newblock \showarticletitle{Correctness of {Sequential} {Monte} {Carlo}
  {Inference} for {Probabilistic} {Programming} {Languages}}. In
  \bibinfo{booktitle}{\emph{Programming {Languages} and {Systems}: 30th
  {European} {Symposium} on {Programming}, {ESOP} 2021, {Held} as {Part} of the
  {European} {Joint} {Conferences} on {Theory} and {Practice} of {Software},
  {ETAPS} 2021, {Luxembourg} {City}, {Luxembourg}, {March} 27 – {April} 1,
  2021, {Proceedings}}}. \bibinfo{publisher}{Springer-Verlag},
  \bibinfo{address}{Berlin, Heidelberg}, \bibinfo{pages}{404--431}.
\newblock
\showISBNx{978-3-030-72018-6}
\urldef\tempurl%
\url{https://doi.org/10.1007/978-3-030-72019-3_15}
\showDOI{\tempurl}


\bibitem[Lundén et~al\mbox{.}(2022)]%
        {lunden22}
\bibfield{author}{\bibinfo{person}{Daniel Lundén}, \bibinfo{person}{Joey
  Öhman}, \bibinfo{person}{Jan Kudlicka}, \bibinfo{person}{Viktor Senderov},
  \bibinfo{person}{Fredrik Ronquist}, {and} \bibinfo{person}{David Broman}.}
  \bibinfo{year}{2022}\natexlab{}.
\newblock \showarticletitle{Compiling {Universal} {Probabilistic} {Programming}
  {Languages} with {Efficient} {Parallel} {Sequential} {Monte} {Carlo}
  {Inference}}. In \bibinfo{booktitle}{\emph{Programming {Languages} and
  {Systems}}} \emph{(\bibinfo{series}{Lecture {Notes} in {Computer}
  {Science}})}, \bibfield{editor}{\bibinfo{person}{Ilya Sergey}} (Ed.).
  \bibinfo{publisher}{Springer International Publishing},
  \bibinfo{address}{Cham}, \bibinfo{pages}{29--56}.
\newblock
\showISBNx{978-3-030-99336-8}
\urldef\tempurl%
\url{https://doi.org/10.1007/978-3-030-99336-8_2}
\showDOI{\tempurl}


\bibitem[Ma et~al\mbox{.}(2021)]%
        {ma21}
\bibfield{author}{\bibinfo{person}{Yingbo Ma}, \bibinfo{person}{Vaibhav Dixit},
  \bibinfo{person}{Michael~J Innes}, \bibinfo{person}{Xingjian Guo}, {and}
  \bibinfo{person}{Chris Rackauckas}.} \bibinfo{year}{2021}\natexlab{}.
\newblock \showarticletitle{A Comparison of Automatic Differentiation and
  Continuous Sensitivity Analysis for Derivatives of Differential Equation
  Solutions}. In \bibinfo{booktitle}{\emph{2021 IEEE High Performance Extreme
  Computing Conference (HPEC)}}. \bibinfo{pages}{1--9}.
\newblock
\urldef\tempurl%
\url{https://doi.org/10.1109/HPEC49654.2021.9622796}
\showDOI{\tempurl}


\bibitem[Matache et~al\mbox{.}(2022)]%
        {matache22}
\bibfield{author}{\bibinfo{person}{Cristina Matache}, \bibinfo{person}{Sean~K.
  Moss}, {and} \bibinfo{person}{Sam Staton}.} \bibinfo{year}{2022}\natexlab{}.
\newblock \showarticletitle{Concrete categories and higher-order recursion:
  With applications including probability, differentiability, and full
  abstraction}. In \bibinfo{booktitle}{\emph{{LICS} '22: 37th Annual {ACM/IEEE}
  Symposium on Logic in Computer Science, Haifa, Israel, August 2 - 5, 2022}},
  \bibfield{editor}{\bibinfo{person}{Christel Baier} {and}
  \bibinfo{person}{Dana Fisman}} (Eds.). \bibinfo{publisher}{{ACM}},
  \bibinfo{pages}{57:1--57:14}.
\newblock
\urldef\tempurl%
\url{https://doi.org/10.1145/3531130.3533370}
\showDOI{\tempurl}


\bibitem[Matsuda(2020)]%
        {matsuda20}
\bibfield{author}{\bibinfo{person}{Kazutaka Matsuda}.}
  \bibinfo{year}{2020}\natexlab{}.
\newblock \showarticletitle{Modular {Inference} of {Linear} {Types} for
  {Multiplicity}-{Annotated} {Arrows}}. In
  \bibinfo{booktitle}{\emph{Programming {Languages} and {Systems}}},
  \bibfield{editor}{\bibinfo{person}{Peter Müller}} (Ed.).
  \bibinfo{publisher}{Springer International Publishing},
  \bibinfo{address}{Cham}, \bibinfo{pages}{456--483}.
\newblock
\showISBNx{978-3-030-44914-8}
\urldef\tempurl%
\url{https://doi.org/10.1007/978-3-030-44914-8_17}
\showDOI{\tempurl}


\bibitem[Mazza and Pagani(2021)]%
        {mazza21}
\bibfield{author}{\bibinfo{person}{Damiano Mazza} {and}
  \bibinfo{person}{Michele Pagani}.} \bibinfo{year}{2021}\natexlab{}.
\newblock \showarticletitle{Automatic differentiation in {PCF}}.
\newblock \bibinfo{journal}{\emph{Proceedings of the ACM on Programming
  Languages}} \bibinfo{volume}{5}, \bibinfo{number}{{POPL}}
  (\bibinfo{year}{2021}), \bibinfo{pages}{1--27}.
\newblock
\urldef\tempurl%
\url{https://doi.org/10.1145/3434309}
\showDOI{\tempurl}


\bibitem[Michel et~al\mbox{.}(2024)]%
        {michel24}
\bibfield{author}{\bibinfo{person}{Jesse Michel}, \bibinfo{person}{Kevin Mu},
  \bibinfo{person}{Xuanda Yang}, \bibinfo{person}{Sai~Praveen Bangaru},
  \bibinfo{person}{Elias~Rojas Collins}, \bibinfo{person}{Gilbert Bernstein},
  \bibinfo{person}{Jonathan Ragan-Kelley}, \bibinfo{person}{Michael Carbin},
  {and} \bibinfo{person}{Tzu-Mao Li}.} \bibinfo{year}{2024}\natexlab{}.
\newblock \showarticletitle{Distributions for {Compositionally}
  {Differentiating} {Parametric} {Discontinuities}}.
\newblock \bibinfo{journal}{\emph{Proc. ACM Program. Lang.}}
  \bibinfo{volume}{8}, \bibinfo{number}{OOPSLA1} (\bibinfo{date}{April}
  \bibinfo{year}{2024}), \bibinfo{pages}{126:893--126:922}.
\newblock
\urldef\tempurl%
\url{https://doi.org/10.1145/3649843}
\showDOI{\tempurl}


\bibitem[Moggi(1991)]%
        {moggi91}
\bibfield{author}{\bibinfo{person}{Eugenio Moggi}.}
  \bibinfo{year}{1991}\natexlab{}.
\newblock \showarticletitle{Notions of Computation and Monads}.
\newblock \bibinfo{journal}{\emph{Information and Computation}}
  \bibinfo{volume}{93}, \bibinfo{number}{1} (\bibinfo{year}{1991}),
  \bibinfo{pages}{55--92}.
\newblock
\urldef\tempurl%
\url{https://doi.org/10.1016/0890-5401(91)90052-4}
\showDOI{\tempurl}


\bibitem[Murray and Schön(2018)]%
        {murray18}
\bibfield{author}{\bibinfo{person}{Lawrence~M. Murray} {and}
  \bibinfo{person}{Thomas~B. Schön}.} \bibinfo{year}{2018}\natexlab{}.
\newblock \showarticletitle{Automated learning with a probabilistic programming
  language: {Birch}}.
\newblock \bibinfo{journal}{\emph{Annual Reviews in Control}}
  \bibinfo{volume}{46} (\bibinfo{date}{Jan.} \bibinfo{year}{2018}),
  \bibinfo{pages}{29--43}.
\newblock
\showISSN{1367-5788}
\urldef\tempurl%
\url{https://doi.org/10.1016/j.arcontrol.2018.10.013}
\showDOI{\tempurl}


\bibitem[Neckel et~al\mbox{.}(2017)]%
        {neckel17}
\bibfield{author}{\bibinfo{person}{Tobias Neckel},
  \bibinfo{person}{Alfredo~Parra Hinojosa}, {and} \bibinfo{person}{Florian
  Rupp}.} \bibinfo{year}{2017}\natexlab{}.
\newblock \bibinfo{title}{Path-Wise Algorithms for Random \& Stochastic ODEs
  with Applications to Ground-Motion-Induced Excitations of Multi-Storey
  Buildings}.  (\bibinfo{year}{2017}).
\newblock
\newblock
\shownote{Unpublished manuscript}.


\bibitem[Pantelides(1988)]%
        {pantelides88}
\bibfield{author}{\bibinfo{person}{Constantinos~C Pantelides}.}
  \bibinfo{year}{1988}\natexlab{}.
\newblock \showarticletitle{The consistent initialization of
  differential-algebraic systems}.
\newblock \bibinfo{journal}{\emph{SIAM J. Sci. Statist. Comput.}}
  \bibinfo{volume}{9}, \bibinfo{number}{2} (\bibinfo{year}{1988}),
  \bibinfo{pages}{213–231}.
\newblock
\urldef\tempurl%
\url{https://doi.org/10.1137/0909014}
\showDOI{\tempurl}


\bibitem[Paquet and Staton(2021)]%
        {paquet21}
\bibfield{author}{\bibinfo{person}{Hugo Paquet} {and} \bibinfo{person}{Sam
  Staton}.} \bibinfo{year}{2021}\natexlab{}.
\newblock \showarticletitle{Lazy{PPL}: laziness and types in non-parametric
  probabilistic programs}. In \bibinfo{booktitle}{\emph{Advances in Programming
  Languages and Neurosymbolic Systems Workshop}}.
\newblock
\urldef\tempurl%
\url{https://openreview.net/forum?id=yHox9OyegeX}
\showURL{%
\tempurl}


\bibitem[Pearlmutter and Siskind(2008)]%
        {pearlmutter08}
\bibfield{author}{\bibinfo{person}{Barak~A. Pearlmutter} {and}
  \bibinfo{person}{Jeffrey~Mark Siskind}.} \bibinfo{year}{2008}\natexlab{}.
\newblock \showarticletitle{Reverse-mode {AD} in a functional framework:
  {Lambda} the ultimate backpropagator}.
\newblock \bibinfo{journal}{\emph{ACM Trans. Program. Lang. Syst.}}
  \bibinfo{volume}{30}, \bibinfo{number}{2} (\bibinfo{date}{March}
  \bibinfo{year}{2008}), \bibinfo{pages}{7:1--7:36}.
\newblock
\showISSN{0164-0925}
\urldef\tempurl%
\url{https://doi.org/10.1145/1330017.1330018}
\showDOI{\tempurl}


\bibitem[Petricek et~al\mbox{.}(2014)]%
        {petricek14}
\bibfield{author}{\bibinfo{person}{Tomas Petricek}, \bibinfo{person}{Dominic
  Orchard}, {and} \bibinfo{person}{Alan Mycroft}.}
  \bibinfo{year}{2014}\natexlab{}.
\newblock \showarticletitle{Coeffects: a calculus of context-dependent
  computation}. In \bibinfo{booktitle}{\emph{Proceedings of the 19th {ACM}
  {SIGPLAN} international conference on {Functional} programming}}
  \emph{(\bibinfo{series}{{ICFP} '14})}. \bibinfo{publisher}{Association for
  Computing Machinery}, \bibinfo{address}{New York, NY, USA},
  \bibinfo{pages}{123--135}.
\newblock
\showISBNx{978-1-4503-2873-9}
\urldef\tempurl%
\url{https://doi.org/10.1145/2628136.2628160}
\showDOI{\tempurl}


\bibitem[Petzold(1982)]%
        {petzold82}
\bibfield{author}{\bibinfo{person}{Linda Petzold}.}
  \bibinfo{year}{1982}\natexlab{}.
\newblock \showarticletitle{Differential/algebraic equations are not ODE’s}.
\newblock \bibinfo{journal}{\emph{SIAM J. Sci. Statist. Comput.}}
  \bibinfo{volume}{3}, \bibinfo{number}{3} (\bibinfo{year}{1982}),
  \bibinfo{pages}{367--384}.
\newblock


\bibitem[Pierce(2002)]%
        {pierce02subtyping}
\bibfield{author}{\bibinfo{person}{Benjamin~C. Pierce}.}
  \bibinfo{year}{2002}\natexlab{}.
\newblock \bibinfo{booktitle}{\emph{Metatheory of Subtyping}}.
\newblock \bibinfo{pages}{209--220}.
\newblock


\bibitem[Pitts(2023)]%
        {pitts23}
\bibfield{author}{\bibinfo{person}{Andrew~M. Pitts}.}
  \bibinfo{year}{2023}\natexlab{}.
\newblock \showarticletitle{Locally Nameless Sets}.
\newblock \bibinfo{journal}{\emph{Proceedings of the ACM on Programming
  Languages}} \bibinfo{volume}{7}, \bibinfo{number}{POPL}, Article
  \bibinfo{articleno}{17} (\bibinfo{date}{jan} \bibinfo{year}{2023}),
  \bibinfo{numpages}{27}~pages.
\newblock
\urldef\tempurl%
\url{https://doi.org/10.1145/3571210}
\showDOI{\tempurl}


\bibitem[Pryce(2001)]%
        {pryce01}
\bibfield{author}{\bibinfo{person}{John~D Pryce}.}
  \bibinfo{year}{2001}\natexlab{}.
\newblock \showarticletitle{A simple structural analysis method for DAEs}.
\newblock \bibinfo{journal}{\emph{BIT Numerical Mathematics}}
  \bibinfo{volume}{41} (\bibinfo{year}{2001}), \bibinfo{pages}{364--394}.
\newblock


\bibitem[Rainforth et~al\mbox{.}(2018)]%
        {rainforth18}
\bibfield{author}{\bibinfo{person}{Tom Rainforth}, \bibinfo{person}{Rob
  Cornish}, \bibinfo{person}{Hongseok Yang}, \bibinfo{person}{Andrew
  Warrington}, {and} \bibinfo{person}{Frank Wood}.}
  \bibinfo{year}{2018}\natexlab{}.
\newblock \showarticletitle{On Nesting {M}onte {C}arlo Estimators}. In
  \bibinfo{booktitle}{\emph{Proceedings of the 35th International Conference on
  Machine Learning}} \emph{(\bibinfo{series}{Proceedings of Machine Learning
  Research}, Vol.~\bibinfo{volume}{80})},
  \bibfield{editor}{\bibinfo{person}{Jennifer Dy} {and}
  \bibinfo{person}{Andreas Krause}} (Eds.). \bibinfo{publisher}{PMLR},
  \bibinfo{pages}{4267--4276}.
\newblock
\urldef\tempurl%
\url{https://proceedings.mlr.press/v80/rainforth18a.html}
\showURL{%
\tempurl}


\bibitem[Rumelhart et~al\mbox{.}(1986)]%
        {rumelhart1986learning}
\bibfield{author}{\bibinfo{person}{David~E Rumelhart},
  \bibinfo{person}{Geoffrey~E Hinton}, {and} \bibinfo{person}{Ronald~J
  Williams}.} \bibinfo{year}{1986}\natexlab{}.
\newblock \showarticletitle{{Learning representations by back-propagating
  errors}}.
\newblock \bibinfo{journal}{\emph{{Nature}}} \bibinfo{volume}{323},
  \bibinfo{number}{6088} (\bibinfo{year}{1986}), \bibinfo{pages}{533--536}.
\newblock


\bibitem[{\'{S}}cibior et~al\mbox{.}(2018)]%
        {scibior18}
\bibfield{author}{\bibinfo{person}{Adam {\'{S}}cibior}, \bibinfo{person}{Ohad
  Kammar}, \bibinfo{person}{Matthijs V{\'{a}}k{\'{a}}r}, \bibinfo{person}{Sam
  Staton}, \bibinfo{person}{Hongseok Yang}, \bibinfo{person}{Yufei Cai},
  \bibinfo{person}{Klaus Ostermann}, \bibinfo{person}{Sean~K. Moss},
  \bibinfo{person}{Chris Heunen}, {and} \bibinfo{person}{Zoubin Ghahramani}.}
  \bibinfo{year}{2018}\natexlab{}.
\newblock \showarticletitle{Denotational validation of higher-order Bayesian
  inference}.
\newblock \bibinfo{journal}{\emph{Proceedings of the ACM on Programming
  Languages}} \bibinfo{volume}{2}, \bibinfo{number}{{POPL}}
  (\bibinfo{year}{2018}), \bibinfo{pages}{60:1--60:29}.
\newblock
\urldef\tempurl%
\url{https://doi.org/10.1145/3158148}
\showDOI{\tempurl}


\bibitem[Shaikhha et~al\mbox{.}(2019)]%
        {shaikhha19}
\bibfield{author}{\bibinfo{person}{Amir Shaikhha}, \bibinfo{person}{Andrew
  Fitzgibbon}, \bibinfo{person}{Dimitrios Vytiniotis}, {and}
  \bibinfo{person}{Simon Peyton~Jones}.} \bibinfo{year}{2019}\natexlab{}.
\newblock \showarticletitle{Efficient differentiable programming in a
  functional array-processing language}.
\newblock \bibinfo{journal}{\emph{Proceedings of the ACM on Programming
  Languages}} \bibinfo{volume}{3}, \bibinfo{number}{ICFP} (\bibinfo{date}{July}
  \bibinfo{year}{2019}), \bibinfo{pages}{97:1--97:30}.
\newblock
\urldef\tempurl%
\url{https://doi.org/10.1145/3341701}
\showDOI{\tempurl}


\bibitem[Sherman et~al\mbox{.}(2021)]%
        {sherman21}
\bibfield{author}{\bibinfo{person}{Benjamin Sherman}, \bibinfo{person}{Jesse
  Michel}, {and} \bibinfo{person}{Michael Carbin}.}
  \bibinfo{year}{2021}\natexlab{}.
\newblock \showarticletitle{$\lambda_s$: computable semantics for
  differentiable programming with higher-order functions and datatypes}.
\newblock  \bibinfo{volume}{5}, \bibinfo{number}{POPL} (\bibinfo{date}{Jan.}
  \bibinfo{year}{2021}), \bibinfo{pages}{3:1--3:31}.
\newblock
\urldef\tempurl%
\url{https://doi.org/10.1145/3434284}
\showDOI{\tempurl}
\newblock
\shownote{Supplement: Implementation of $\lambda_s$}.


\bibitem[Siskind and Pearlmutter(2005)]%
        {siskind05}
\bibfield{author}{\bibinfo{person}{Jeffrey~Mark Siskind} {and}
  \bibinfo{person}{Barak~A Pearlmutter}.} \bibinfo{year}{2005}\natexlab{}.
\newblock \bibinfo{title}{Perturbation confusion and referential transparency:
  Correct functional implementation of forward-mode AD}.
  (\bibinfo{year}{2005}).
\newblock
\newblock
\shownote{Unpublished manuscript, presented at the 17th International Workshop
  on Implementation and Application of Functional Languages (IFL2005), Sep
  19-21, 2005, Dublin Ireland.}.


\bibitem[Siskind and Pearlmutter(2008)]%
        {siskind08}
\bibfield{author}{\bibinfo{person}{Jeffrey~Mark Siskind} {and}
  \bibinfo{person}{Barak~A. Pearlmutter}.} \bibinfo{year}{2008}\natexlab{}.
\newblock \showarticletitle{Nesting forward-mode {AD} in a functional
  framework}.
\newblock \bibinfo{journal}{\emph{Higher-Order and Symbolic Computation}}
  \bibinfo{volume}{21}, \bibinfo{number}{4} (\bibinfo{date}{Dec.}
  \bibinfo{year}{2008}), \bibinfo{pages}{361–376}.
\newblock
\showISSN{1573-0557}
\urldef\tempurl%
\url{https://doi.org/10.1007/s10990-008-9037-1}
\showDOI{\tempurl}


\bibitem[V{\'{a}}k{\'{a}}r et~al\mbox{.}(2019)]%
        {vakar19}
\bibfield{author}{\bibinfo{person}{Matthijs V{\'{a}}k{\'{a}}r},
  \bibinfo{person}{Ohad Kammar}, {and} \bibinfo{person}{Sam Staton}.}
  \bibinfo{year}{2019}\natexlab{}.
\newblock \showarticletitle{A domain theory for statistical probabilistic
  programming}.
\newblock \bibinfo{journal}{\emph{Proceedings of the ACM on Programming
  Languages}} \bibinfo{volume}{3}, \bibinfo{number}{{POPL}}
  (\bibinfo{year}{2019}), \bibinfo{pages}{36:1--36:29}.
\newblock
\urldef\tempurl%
\url{https://doi.org/10.1145/3290349}
\showDOI{\tempurl}


\bibitem[Wang et~al\mbox{.}(2019)]%
        {wang19}
\bibfield{author}{\bibinfo{person}{Fei Wang}, \bibinfo{person}{Daniel Zheng},
  \bibinfo{person}{James Decker}, \bibinfo{person}{Xilun Wu},
  \bibinfo{person}{Grégory~M. Essertel}, {and} \bibinfo{person}{Tiark Rompf}.}
  \bibinfo{year}{2019}\natexlab{}.
\newblock \showarticletitle{Demystifying differentiable programming:
  shift/reset the penultimate backpropagator}.
\newblock \bibinfo{journal}{\emph{Proceedings of the ACM on Programming
  Languages}} \bibinfo{volume}{3}, \bibinfo{number}{ICFP} (\bibinfo{date}{July}
  \bibinfo{year}{2019}), \bibinfo{pages}{1--31}.
\newblock
\showISSN{2475-1421}
\urldef\tempurl%
\url{https://doi.org/10.1145/3341700}
\showDOI{\tempurl}


\bibitem[Wood et~al\mbox{.}(2014)]%
        {wood14}
\bibfield{author}{\bibinfo{person}{Frank Wood}, \bibinfo{person}{Jan~Willem
  Meent}, {and} \bibinfo{person}{Vikash Mansinghka}.}
  \bibinfo{year}{2014}\natexlab{}.
\newblock \showarticletitle{{A New Approach to Probabilistic Programming
  Inference}}. In \bibinfo{booktitle}{\emph{Proceedings of the Seventeenth
  International Conference on Artificial Intelligence and Statistics}}
  \emph{(\bibinfo{series}{Proceedings of Machine Learning Research},
  Vol.~\bibinfo{volume}{33})}, \bibfield{editor}{\bibinfo{person}{Samuel Kaski}
  {and} \bibinfo{person}{Jukka Corander}} (Eds.). \bibinfo{publisher}{PMLR},
  \bibinfo{address}{Reykjavik, Iceland}, \bibinfo{pages}{1024--1032}.
\newblock
\urldef\tempurl%
\url{https://proceedings.mlr.press/v33/wood14.html}
\showURL{%
\tempurl}


\end{thebibliography}

\pagebreak

%%
%% If your work has an appendix, this is the place to put it.
\appendix

\section{Lotka-Volterra Model}\label{apdx:lotak-volterra}

The Lotka-Volterra predator-prey model is defined as the following coupled ODEs:

\begin{equation}
	\begin{cases}
		\frac{d}{dx}y_1 & = \theta_1 y_1 - \theta_2y_1y_2    \\
		\frac{d}{dx}y_2 & = \theta_3 y_1 y_1 - \theta_4 y_1,
	\end{cases}
\end{equation}
where \(y = y(x)\); \(y_1\) and \(y_2\) are prey and predator densities, respectively; \(\theta_1\) and \(\theta_3\) are the per-capita growth rate of prey and predators, respectively; \(\theta_2\) is the prey death rate from predators; and \(\theta_4\) is the predator death rate from a lack of prey.
\end{document}